\documentclass[sigconf]{acmart}
\AtBeginDocument{%
  }

\setcopyright{acmlicensed}
\copyrightyear{2025}
\acmYear{2025}
\acmDOI{10.1145/3746027.3755466}
\acmConference[MM '25] {Proceedings of the 33rd ACM International Conference on Multimedia}{October 27--31, 2025}{Dublin, Ireland.}
\acmBooktitle{Proceedings of the 33rd ACM International Conference on Multimedia (MM '25), October 27--31, 2025, Dublin, Ireland}
\acmISBN{979-8-4007-2035-2/2025/10}

\usepackage{booktabs}
\usepackage{siunitx}
\usepackage{multirow}
\usepackage{subcaption}
\usepackage{rotating}
\usepackage[skip=0.5ex]{caption}
\usepackage{geometry}
\usepackage[table]{xcolor}
\definecolor{best}{RGB}{217,234,211}  
\definecolor{second}{RGB}{254,249,205} 
\usepackage[misc]{ifsym}

\begin{document}

\title{Regist3R: Incremental Registration with Stereo Foundation Model}

\author{Sidun Liu}
\affiliation{
  \institution{College of Computer Science and Technology\\
  National Key Laboratory of Parallel and Distributed Computing\\
  National University of Defense Technology}
  \city{Changsha}
  \state{Hunan}
  \country{China}
}
\email{liusidun@nudt.edu.cn}
\author{Wenyu Li}
\affiliation{
  \institution{College of Computer Science and Technology\\
  National Key Laboratory of Parallel and Distributed Computing\\
  National University of Defense Technology}
  \city{Changsha}
  \state{Hunan}
  \country{China}
}
\email{wenyu18@nudt.edu.cn}
\author{Peng Qiao}
\authornotemark[1]
\affiliation{
  \institution{College of Computer Science and Technology\\
  National Key Laboratory of Parallel and Distributed Computing\\
  National University of Defense Technology}
  \city{Changsha}
  \state{Hunan}
  \country{China}
}
\email{pengqiao@nudt.edu.cn}
\author{Yong Dou}
\authornote{Peng Qiao and Yong Dou are Corresponding authors.}
\affiliation{
  \institution{College of Computer Science and Technology\\
  National Key Laboratory of Parallel and Distributed Computing\\
  National University of Defense Technology}
  \city{Changsha}
  \state{Hunan}
  \country{China}
}
\email{yongdou@nudt.edu.cn}

\renewcommand{\shortauthors}{Sidun Liu, Wenyu Li, Peng Qiao, and Yong Dou}

\begin{abstract}
  Multi-view 3D reconstruction has remained an essential yet challenging problem in the field of computer vision. While DUSt3R and its successors have achieved breakthroughs in 3D reconstruction from unposed images, these methods exhibit significant limitations when scaling to multi-view scenarios, including high computational cost and cumulative error induced by global alignment. 
  To address these challenges, we propose Regist3R, a novel stereo foundation model tailored for efficient and scalable incremental reconstruction. Regist3R leverages an incremental reconstruction paradigm, enabling large-scale 3D reconstructions from unordered and many-view image collections. 
  We evaluate Regist3R on public datasets for camera pose estimation and 3D reconstruction.
  Our experiments demonstrate that Regist3R achieves comparable performance with optimization-based methods while significantly improving computational efficiency, and outperforms existing multi-view reconstruction models.
  Furthermore, to assess its performance in real-world applications, we introduce a challenging oblique aerial dataset which has long spatial spans and hundreds of views. The results highlight the effectiveness of Regist3R. We also demonstrate the first attempt to reconstruct large-scale scenes encompassing over thousands of views through pointmap-based foundation models, showcasing its potential for practical applications in large-scale 3D reconstruction tasks, including urban modeling, aerial mapping, and beyond. 
\end{abstract}





\maketitle


\begin{figure*}[t]
    \includegraphics[clip,trim=1mm 1mm 1mm 1mm, width=\linewidth]{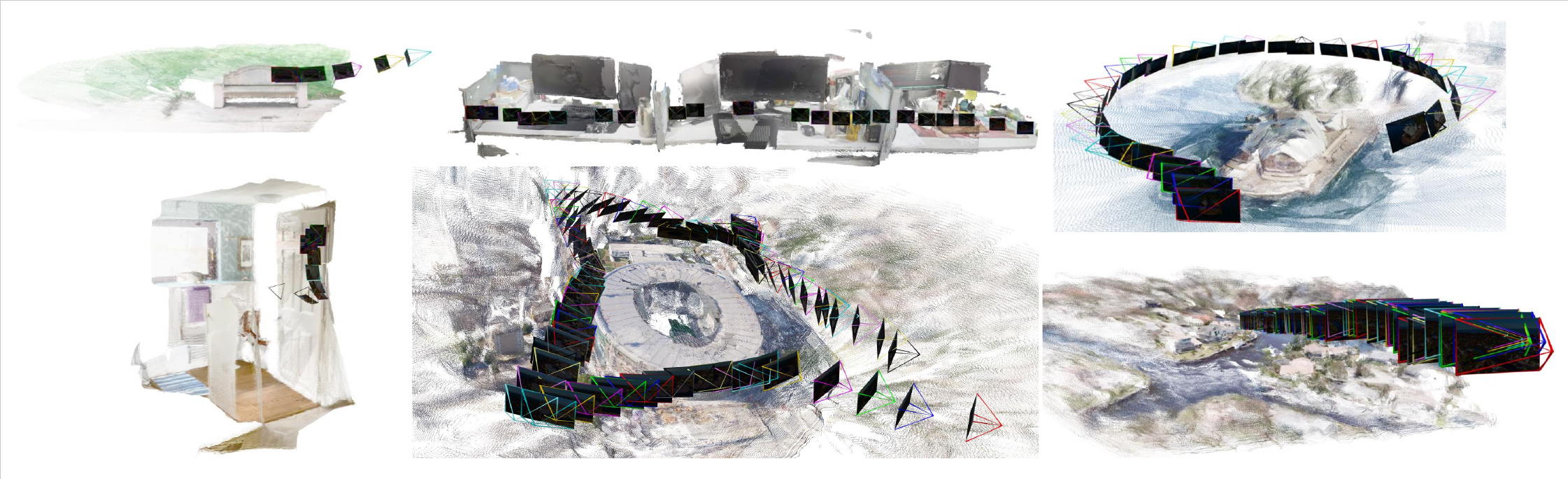}
    \caption{Qualitative examples of Regist3R, covering indoor, outdoor, and aerial scenes, demonstrating its generalizability.}
    \label{fig:examples}
   \end{figure*}
   
\section{Introduction}


Multi-view 3D reconstruction is a fundamental task in computer vision, enabling the reconstruction of 3D models from multiple 2D images. The challenge lies in accurately estimating camera poses and reconstructing 3D structures from these images, particularly in complex environments with large spatial spans.

As the cornerstone of multi-view 3D reconstruction, Structure from Motion (SfM)~\cite{agarwal2011building} reconstructs 3D structures and camera poses from unordered 2D images, but struggles with feature matching noise, scale drift, and scalability. Although global SfM~\cite{moulon2013global,cai2021pose,wilson2014robust} optimizes all camera parameters simultaneously, its dependency on reliable initial pairwise geometries limits its effectiveness in scenes with sparse features. Incremental SfM~\cite{schonberger2016structure,snavely2006photo,wu2013towards} mitigates these issues through sequential image registration and iterative refinement, proving to be more resilient in unstructured environments where global methods often fail. This adaptability explains its prevalent use in real-world applications with imperfect data conditions.

Traditional SfM pipelines rely on sequential modules—feature extraction, matching, relative pose estimation, and triangulation~\cite{lowe2004distinctive,sarlin2020superglue}—a fragmented paradigm susceptible to cumulative error propagation. Recent learning-based methods~\cite{wang2024vggsfm,brachmann2024scene,wang2024dust3r} attempt to replace these handcrafted components with end-to-end formulations. While VGGSfM~\cite{wang2024vggsfm} introduces differentiable submodules and Detector-free SfM~\cite{he2024detector} employs learned feature matching, both retain the classical pipeline structure. More radical approaches like FlowMap~\cite{smith2024flowmap} and Ace-Zero~\cite{brachmann2024scene} abandon modular designs entirely, directly optimizing geometry via per-scene gradient descent with geometry regressor networks. However, such methods remain constrained to scenarios with high image overlap and stable illumination. 

In this landscape, DUSt3R~\cite{wang2024dust3r} emerges as a breakthrough by predicting globally consistent pointmaps in a unified coordinate system from two views, demonstrating exceptional geometric coherence without incremental error accumulation. Building on this success, recent extensions~\cite{leroy2024grounding,duisterhof2024mast3r,yang2025fast3r,tang2024mv,wang2025vggt,elflein2025light3r,cabon2025must3r,wang20243d} explore multi-view generalizations of DUSt3R's framework, which demonstrate divergent optimization paradigms. Post-optimization approaches, exemplified by MASt3R-SfM~\cite{duisterhof2024mast3r} and Light3R-SfM~\cite{elflein2025light3r}, iteratively reconstruct scenes from pairwise matchings: MASt3R-SfM leverages pointmaps and matches from MASt3R for bundle adjustment, while Light3R-SfM accelerates convergence via direct Procrustes alignment~\cite{umeyama1991least}. However, these methods fail to establish fully differentiable reconstruction pipelines driven by Transformer inference. Alternatively, another research direction~\cite{yang2025fast3r,tang2024mv} explores direct multi-view geometry prediction through scaled Transformer architectures. By integrating cross-attention mechanisms, these models enforce coordinate consistency across views during single-forward passes. Yet their scalability remains fundamentally constrained by the quadratic complexity growth of cross-attention layers with increasing view numbers. These limitations expose a critical gap in current learning-based SfM systems—achieving both optimization-free and scalable multi-view generalization.

Multi-view 3D reconstruction models often suffer from limited generalization when extrapolated beyond their training distribution in viewpoint diversity or scene scale, whereas stereo models achieve stronger robustness by leveraging low-level geometric primitives that inherently constrain spatial reasoning across extended viewpoints and environments.
In this paper, we try to answer a challenging question: \textit{Can we build an inference-only and scalable SfM system with just Stereo Foundation Models?}


To address this challenge, we present Regist3R, a stereo foundational model for incremental registration. Regist3R is a transformer-based~\cite{vaswani2017attention} network that regresses the pointmap of an unregistered view in the world coordinate system by leveraging a given reference view along with its world coordinate pointmap. An intuitive formulation and illustration compared to DUSt3R is shown as Eq.~\ref{eq:formulation} and Fig.~\ref{fig:architecture},
\begin{equation}
 \label{eq:formulation}
 \begin{aligned}
 I^1 + I^2 &\xrightarrow{DUSt3R} X^{1,1} + X^{2,1} \\
 I^i + I^j + X^{i,1} &\xrightarrow{Regist3R} X^{j,1}.
 \end{aligned}
\end{equation}
Analogous to DUSt3R, Regist3R uses a two-stream architecture with cross-attention to encode reference pointmap and target image respectively. Only one regression head is kept for the estimation of target pointmap. Although the model only accepts stereo input, it can be scaled to any size of viewpoints via autoregression. 

Drifting is one of the most critical problems for incremental reconstruction~\cite{holynski2020reducing}, and time-consuming bundle adjustment (BA) is adopted by most of traditional pipelines~\cite{schonberger2016structure}. In the absence of BA for inference-based models, we enhance Regist3R's noise resistance to mitigate drift. Specifically, in addition to feeding the model ground truth pointmap as input, we use a chain training strategy to estimate target structure based on inaccurate pointmap.

For efficient inference, we build a minimum spanning tree (MST) based on the view similarity, and perform pairwise inference only between the parents and children. This takes only $N-1$ times inference for a collection containing $N$ images. As the drifting happens when the tree goes deep, we adopt a tree compression trick to reduce the length of reconstruction chain.

We summarize the key contributions of this work as follows:

\begin{itemize}
 \item We propose Regist3R, a stereo foundation model for inference-only and scalable incremental registration. Based on our Regist3R, we further build a feed-forward incremental SfM pipeline for efficient reconstruction.
 \item The experiments show that the performance of Regist3R is on par with or even better than optimization-based approaches and multi-view models. Furthermore, Regist3R demonstrates significant superiority over comparable methods when the scene has more views and larger spatial span.
\end{itemize}

\section{Related Works}

\subsection{Structure-from-Motion}

Structure-from-Motion (SfM)~\cite{agarwal2011building} is a pivotal technique in computer vision for reconstructing 3D scenes from 2D images. Traditional SfM pipelines predominantly employ two approaches: global and incremental methods.

Global SfM~\cite{jiang2013global,moulon2013global,wilson2014robust,ozyesil2015robust,moulon2016openmvg} addresses the entire dataset simultaneously, estimating camera positions and 3D structures in a unified optimization process. While this approach offers efficiency, it often faces challenges related to scalability and precision. The computational demands can be substantial, especially with large datasets, and maintaining accuracy across extensive reconstructions remains a concern. Recent advancements, such as the GLOMAP~\cite{pan2024global} and XM~\cite{han2025building}, aim to enhance global SfM by balancing efficiency with improved accuracy, making it more viable for complex modeling tasks.

Incremental SfM~\cite{snavely2008modeling,wu2013towards,schonberger2016structure,mur2017orb}, on the other hand, reconstructs scenes by progressively adding images, starting from an initial pair and iteratively integrating new ones. This method is renowned for its accuracy and robustness, particularly in well-textured environments. However, its time complexity is often considered $O(n^4)$ concerning the number of images, leading to inefficiencies as dataset sizes grow. Hybrid SfM~\cite{cui2017hsfm,zhu2017parallel,ye2024er,ye2024absgs} approaches attempt to combine both for balance between efficiency and scalability.

Traditional SfM methods, particularly those based on optimization, often encounter limitations in both efficiency and accuracy. The computational burden of global optimization can be prohibitive, and the incremental nature of certain methods may not scale well with increasing data volume. Although learning-based methods~\cite{sarlin2020superglue,lindenberger2021pixel,dusmanu2019d2,gleize2023silk,teed2021droid,yi2016lift,wang2023posediffusion,li2025sfmdiffusion,zhang2024cameras} replace some steps, the cumulative error of the pipeline still limits its scalability. In contrast, our model directly learns the incremental registration of SfM, eliminating the cumulative error caused by the pipeline. The inference-only mode eliminates the need for optimization and improves registration efficiency.

\subsection{Reconstruction in Large Model Era}

Instead of replacing some steps of the SfM pipeline, recent works have proposed to learn the reconstruction task in an end-to-end manner~\cite{zhao2022particlesfm,wang2024vggsfm,brachmann2024scene}. VGGSfM~\cite{wang2024vggsfm} makes individual SfM components learnable forming a fully differentiable SfM framework. ACEZero\cite{brachmann2024scene} proposes to incrementally optimize scene coordinate regression and camera refinement networks to minimize reprojection errors. Similarly, FlowMap optimizes the per-scene depth estimation network with offline optical flow and point tracking supervision. Yet the performance is limited when the image set shares low visual overlapping.

The recently proposed Dense and Unconstrained Stereo 3D Reconstruction model DUSt3R~\cite{wang2024dust3r} brings a new paradigm to 3D reconstruction. It proposed a novel approach for two-view reconstruction via direct pointmap regression from a pair of RGB images, and the pointmaps of two input views share the same coordinate system. The improved version MASt3R~\cite{leroy2024grounding} further extends DUSt3R with image matching. When handling multi-view reconstruction, the exhaustive pairwise inference constrains the scalability of DUSt3R. MASt3R-SfM~\cite{duisterhof2024mast3r} incorporates image retrieval and builds an optimization pipeline on MASt3R matches. However, the optimization process is time-consuming and the precision relies on matching accuracy. Multiple works extend DUSt3R to multi-view scenarios. Analogous to traditional approaches, we categorize them into global and incremental. Global approaches extend the model to accept multi-view images and use wide cross-attention for information sharing. Fast3R~\cite{yang2025fast3r} uses random index embedding to improve model generalizability. MV-DUSt3R~\cite{tang2024mv} uses cross-reference-view blocks to make it robust to reference view selection. VGGT~\cite{wang2025vggt} proposes to predict multiple non-orthogonal variables, like camera pose, depthmap, and pointmap, meanwhile using alternative attention for global alignment. Light3R-SfM~\cite{elflein2025light3r} adopts a latent global alignment module to reduce the cost of global attention. However, the training cost is high and the scalability is limited by the model generalizability. Incremental approaches estimate the 3D structure of the current view based on previous reconstructed structures. Spann3R~\cite{wang20243d} maintains a memory bank for sequential reconstruction, but the scenario is limited to sequential images, and the generalizability of implicit memory bank representation of history frames is questionable. MUSt3R~\cite{cabon2025must3r} proposes a similar implicit memory mechanism, therefore, its application scenarios are still limited to online reconstruction of sequential images.

Our Regist3R falls into the incremental category. Unlike previous works that use implicit memory banks to encode history features, it directly adopts explicit pointmap representation, which gives it a more flexible reconstruction path and supports offline reconstruction of unordered image collections.


\begin{figure}[t]
    \includegraphics[clip,trim=8mm 8mm 8mm 5mm, width=\linewidth]{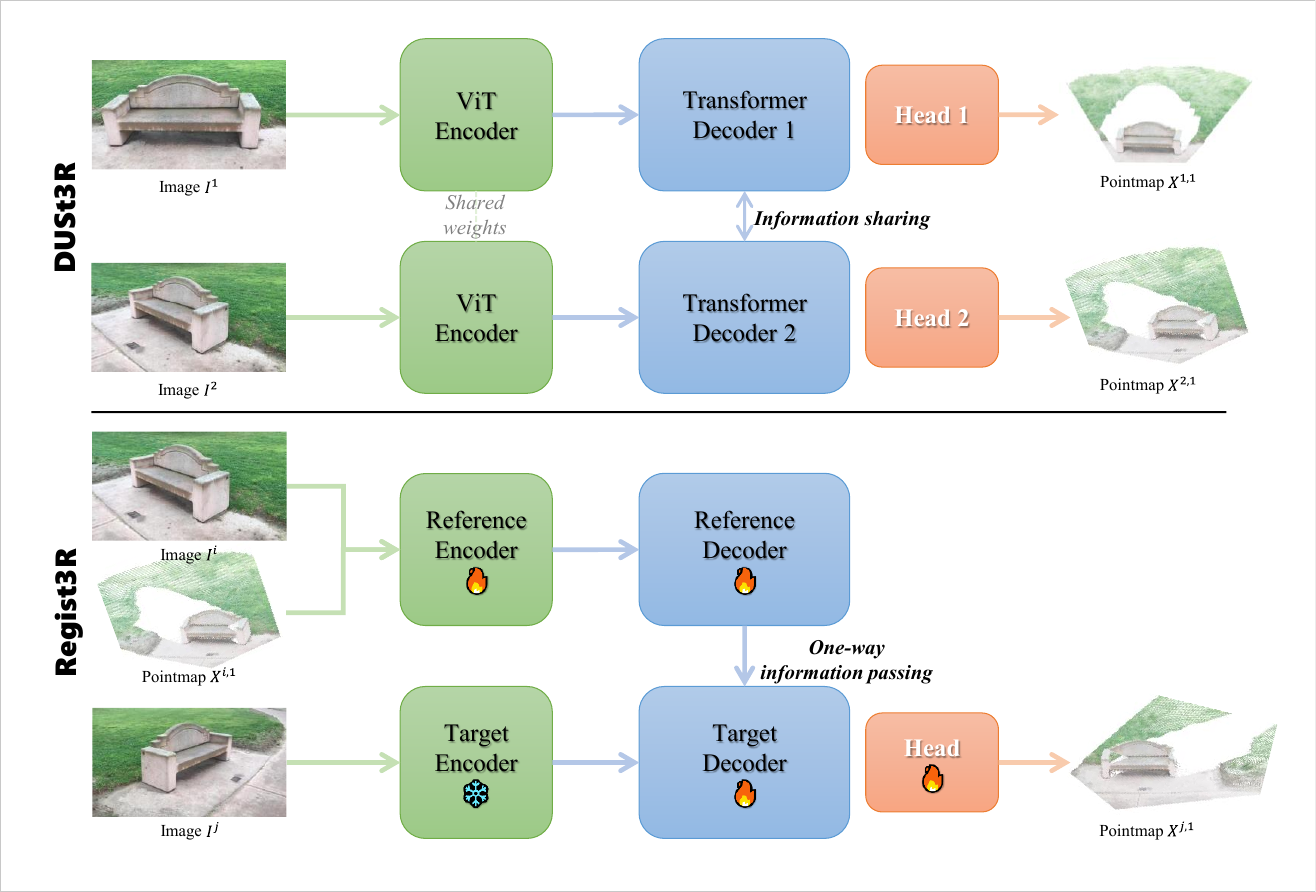}
    \caption{Comparison between Regist3R and DUSt3R. (Top) Two output pointmaps of \textit{DUSt3R} share the same coordinate system, (Bottom) while the input and output pointmaps of \textit{Regist3R} share the same coordinate system. The accompanying confidences are omitted.}
    \label{fig:architecture}
\end{figure}

\section{Approach}


The two-view foundation model DUSt3R~\cite{wang2024dust3r} accepts two images in the same coordinate system and outputs their respective point maps. However, when extending to multiple views, it requires exhaustively traversing all pairs and performing post-optimization. Subsequent multi-view extensions~\cite{tang2024mv,yang2025fast3r,wang2025vggt} are still limited to a relatively small scale of views. Therefore, a crucial question is how to design a foundational model that achieves true scalability. In this section, we analyze and address this issue from the perspectives of problem formulation, model architecture, training procedure, and inference strategy.

\subsection{Problem Formulation}


We illustrate the problem formulation of Regist3R by constrasting it with DUSt3R, as shown in Fig.~\ref{fig:architecture}. DUSt3R takes two images $I^1, I^2 \in \mathbb{R}^{H \times W \times 3}$ as input and outputs two pointmaps $X^{1,1}, X^{2,1} \in \mathbb{R} ^{H \times W \times 3}$ in the coordinate system defined by the first view $I^1$. When handling multi-views, this leads to inconsistency of coordinate systems and requires additional global alignment. 
In contrast, Regist3R takes as input an reference view $I^i \in \mathbb{R}^{H \times W \times 3}$ along with its pointmap $X^i \in\mathbb{R} ^{H \times W \times 3}$, and a target view $I^j \in \mathbb{R}^{H \times W \times 3}$, and output the pointmap $X^j \in \mathbb{R} ^{H \times W \times 3}$ of the target view in the coordinate system defined by $X^i$. In this case the coordinate system of $X^i$ can be chosen arbitrarily, which allows multi-views to share the same coordinate system, thereby bypassing the need for global alignment. All the input and output pointmaps above are accompanied by their confidences in the domain of $\mathbb{R}^{H \times W}$, which are omitted for simplicity.

\subsection{Model Architecture}

Regist3R is a two-stream transformer-based network similar to DUSt3R, containing two image/pointmap encoders, two decoders equipped with cross-attention, and a single regression head for target pointmap estimation, as shown in Fig.~\ref{fig:architecture}. 

Differently from DUSt3R, the two encoders do not share weights, as the reference encoder encodes the concatenation of the image and pointmap while the target encoder only encodes the image:

\begin{equation}
 \label{eq:encoder}
 \begin{aligned}
 F^i &= \text{Encoder}^i(I^i \Vert X^i) \\
 F^j &= \text{Encoder}^j(I^j).
 \end{aligned}
\end{equation}

The target encoder prepares features for the latent feature matching in the next stage, which is consistent with the encoder of DUSt3R, therefore we reuse and freeze its weights. But the function of the reference encoder needs to be extended to encode images and pointmaps simultaneously, so we expand the patch embedding from 3 channels (RGB) to 6 channels (RGB-XYZ). We use the encoder parameters of DUSt3R as initialization, and fine-tune it during training.

In the decoding stage, the information exchange is one-way, from the reference to the target, therefore the reference decoder only performs self-attention while the target decoder performs the cross-attention and self-attention iteratively. The decoders are the stack of several attention blocks. For each block, it attends to the tokens of the previous layer, and the block of the target decoder also attends to the reference tokens:
\begin{equation}
 \label{eq:decoder}
 \begin{aligned}
 G^i_l &= \text{DecoderBlock}^i_l(G^i_{l-1}) \\
 G^j_l &= \text{DecoderBlock}^j_l(G^j_{l-1}, G^i_{l-1}).
 \end{aligned}
\end{equation}
for $l=1,...,B$ for decoder with $B$ blocks and initialized with encoder tokens $G_0^i:=F^i$ and $G_0^j:=F^j$. The target decoder weights are initialized with DUSt3R's first decoder and the reference decoder weights are optimized from scratch.

Finally, the regression head takes the target decoder tokens and outputs the pointmap of the target view, as well as an associated confidence map:
\begin{equation}
 \label{eq:regression}
 X^j, C^j = \text{Head}(G^j_0,...,G^j_B).
\end{equation}

As proved by DUSt3R, two-stream network architecture achieves impressive performance on relative pointmap prediction. Meanwhile, multi-view network architectures~\cite{yang2025fast3r,wang2025vggt} suffer from long sequence cross-attention and generalizability on more views than training set. Therefore, we keep the two-stream architecture for Regist3R, and extend it to multi-view scenarios by autoregression. In Sec.~\ref{sec:inference}, we will demonstrate the advantages of Regist3R's two-stream architecture in long-sequence inference.

\subsection{Training Procedure}

\subsubsection{Pointmap Preperation}

Multi-view pointmaps may exhibit a broader range of coordinate variation. However, neural networks typically demonstrate better predictive performance around zero-centered data distributions. Therefore, pointmaps normalization is required to facilitate model training.

Formally, sampling a pair of images and ground-truth pointmaps $(I^i, \bar{X}^i), (I^j, \bar{X}^j)$, the pointmaps are normalized by $\bar{X}^i$'s mean $\mu(\bar{X}^i)$ and average distance to the mean $z(\bar{X}^i)$, where $\mathcal{N}_i$ is the normalization operator about $\bar{X}^i$:
\begin{equation}
 \label{eq:pointmap_norm}
 \mathcal{N}_i(X) = \frac{X - \mu(\bar{X}^i)}{z(\bar{X}^i)}, \quad z(\bar{X}^i) = \text{norm}\left(\bar{X}^i-\mu(\bar{X}^i)\right).
\end{equation}
We also apply a random rotation $R \sim \text{Uniform}(SO(3))$ on both pointmaps to ensure the reference map is in an arbitrary coordinate system. Regist3R, marked as $f_\theta$, receives the normalized pointmap and pair of images as input, estimating the target pointmap and confidence map:
\begin{equation}
 \label{eq:input}
 X^j, C^j = f_\theta\left[I^i\Vert \mathcal{N}_i\left(R\bar{X}^i\right), I^j\right]
\end{equation}

We follow DUSt3R to use symmetric and confidence-aware regression loss for training, which is defined~\footnote{We ignore per-valid-pixel summation for simplicity.} as:
\begin{equation}
 \label{eq:loss}
 \begin{aligned}
 \mathcal{L}_{\text{regr}}(i, j) &= \left\Vert\mathcal{N}_i\left(R\bar{X}^j\right) - X^j\right\Vert \\
 \mathcal{L}_{\text{conf}}(i, j) &= C^j\mathcal{L}_{\text{regr}}(i, j) - \alpha \log C^j \\
 \mathcal{L} &= \mathcal{L}_{\text{conf}}(i,j) + \mathcal{L}_{\text{conf}}(j,i)
 \end{aligned}
\end{equation}

Based on explicit pointmap representation, we can normalize the pointmap before each inference. This ensures that, even in large-scale scenarios where the coordinate variation is significant, both the input and output pointmaps remain within a reasonable numerical range.

\subsubsection{Autoregressive Training}

\begin{figure}[t]
    \centering
    \includegraphics[clip,trim=8mm 8mm 5mm 5mm, width=\linewidth]{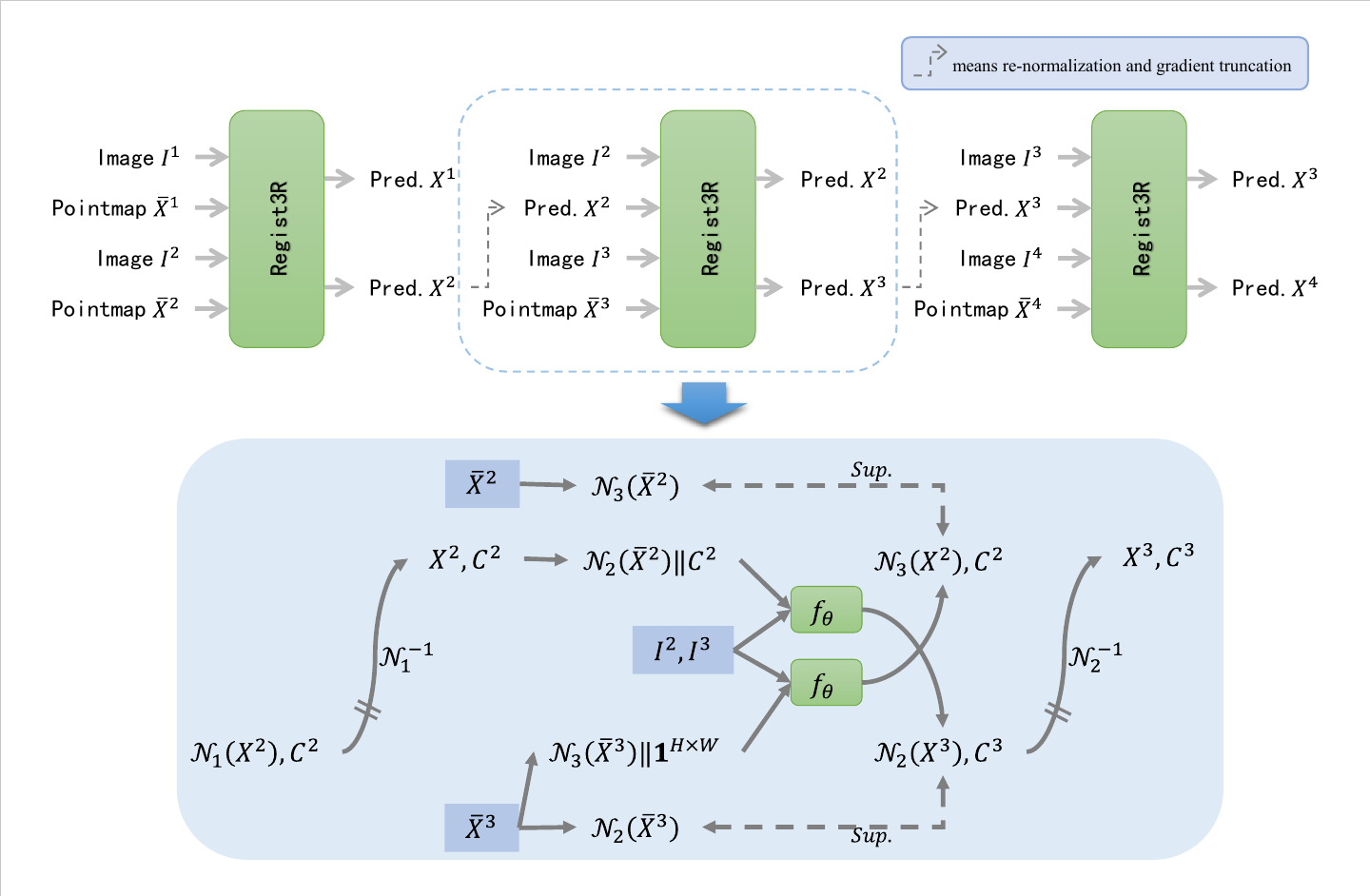}
    \caption{Autoregressive training strategy. }
    \label{fig:chain}
\end{figure}

During incremental registration, the precision of the reference pointmap is not guaranteed and the model should be robust to noise. 
To achieve this, we adopt a chain training strategy, where the model is trained to estimate the target structure based on the inaccurate pointmap and the associated confidence. Sepecifically, we sample a chain of images and ground-truth pointmaps $(I^1, \bar{X}^1),...,(I^N, \bar{X}^N)$. For each step, the previously predicted pointmap and confidence are used as input. As the confidence value is activated by $exp$, we normalize it with \textit{log-sigmoid}, and the confidence of ground truth is all one. Given the output pointmap of previous step $X^n, C^n (n=1,...,N-1)$, and the image pair $I^n, I^{n+1}$, the next step pointmap $X^{n+1}, C^{n+1}$ is estimated by:
\begin{equation}
 \label{eq:chain}
 \begin{aligned}
 X^{n+1}, C^{n+1} = f_\theta\left[I^n\Vert \mathcal{N}_n\mathcal{N}_{n-1}^{-1}X^n\Vert \sigma\circ\log(C^n), I^{n+1}\right]
 \end{aligned}
\end{equation}
The patch embedding of reference encoder is extended to 7 channels (RGB-XYZ-C) to fit confidence input.
Different from Spann3R~\cite{wang20243d} which transfers implicit features across steps and thus requires cross-step optimization, our Regist3R uses pointmap and confidence to transfer information, so each step is trained independently and there is no need to transfer gradients between steps.
A training procedure with a chain length of 3 is shown in Fig.~\ref{fig:chain}.


\subsection{Inference Strategy}
\label{sec:inference}

When inference, previous approaches exhaustively perform pairwise inference on all image pairs~\cite{wang2024dust3r} or a subset of pairs~\cite{duisterhof2024mast3r}, resulting in computational bottleneck. Instead, we follow incremental registration paradigm~\cite{schonberger2016structure}, requiring only one inference for each new view, thus being highly efficient even when the scene scales larger.
Specifically, we build a minimum spanning tree (MST) based on the view similarity, and perform pairwise inference only between the parents and children. This takes only $N-1$ times inference for a collection containing $N$ views. As the drifting happens when the tree goes deep, we propose a tree compression trick to reduce the length of reconstruction chain. 

\subsubsection{Incremental Registration}

For each registration step, traditional approaches, etc. COLMAP~\cite{schonberger2016structure}, select the next images based on the matching with registered images. While matching is not explicitly defined in Regist3R, we resort to image retrieval to select the next image. Following MASt3R-SfM~\cite{duisterhof2024mast3r}, we use Aggregated Selective Matching Kernel~(ASMK)~\cite{tolias2013aggregate} for efficient pairwise similarity measurement, which has shown excellent performance for retrieval. ASMK receives the feature map from the MASt3R~\cite{leroy2024grounding}'s image encoder and outputs a similarity matrix. Then we build a minimum spanning tree (MST) and select the image with the largest summation of similarity with other images as the root~\cite{elflein2025light3r}. As the pointmap of the root is required for bootstrapping, we adopt DUSt3R and select the root and its first child as the initial pair for pointmap estimation. Other approaches, like monocular MoGe~\cite{wang2024moge} or multi-view VGGT~\cite{wang2025vggt}, are also applicable.
Then we set camera coordinate system of root view as the global coordinate system, and travel the MST to estimate the pointmap of each view. 

\subsubsection{Tree Depth Compression}

MST keeps the edge set with minimal cost, when the images are collected in a sequence, the MST tends to degenerate into a chain, which causes the drifting. Light3R-SfM~\cite{elflein2025light3r} proposes to replace MST with a shortest path tree (SPT) to reduce the depth, but it leads to a large disparity between pairs when the scene views are sparse. Inspired by the key frame selection in Spann3R~\cite{wang20243d}, we extend it to tree structure as key layer selection. Specifically, assuming that the key layer selection interval is $K$, the descendant nodes from the $nK+1^{th}$ layer to the $(n+1)K^{th}$ layer select their ancestors at the $nK^{th}$ layer as the reference view. For implementation, we change the parent node of the even-layer node to the grandparent node to achieve $K=2$ compression. Repeating it achieves power-of-2 compression. For dense view, we can set aggressive compression while for sparse view, we can set it smaller.

\subsubsection{Pose Estimation}

In Regist3R, we assume the images share the same intrinsic parameters, and the camera pose is derived from the pointmap. The DUSt3R estimated pointmap is used to derive the camera intrinsics, which is then broadcasted to all other views. Then the camera pose of each view can be derived from the pointmap and the intrinsics with the PnP algorithm.

\textbf{Discussion.} The assumption of sharing intrinsics is necessary for Regist3R, as the global pointmap can't carry enough information to derive both intrinsics and extrinsics. Several approaches simultaneously estimate local and global pointmap, and derive the intrinsics from the local pointmap and the extrinsic from the Procrustes alignment. However, it may suffer from redundancy and misalignment between local and global pointmap. A more refined extension of the pointmap representation needs to be proposed to enable the derivation of both.

\begin{figure*}[ht]
 \centering
 \renewcommand{\arraystretch}{1.2}
 \begin{tabular}{@{}c|cccc@{}}
 & \multicolumn{1}{c}{\textbf{DUSt3R$^\dagger$}} 
 & \multicolumn{1}{c}{\textbf{MASt3R-SfM}} 
 & \multicolumn{1}{c}{\textbf{Regist3R}} 
 & \multicolumn{1}{c}{\textbf{GT Poses}} \\[1ex]
 \hline
 
 \raisebox{1cm}[0pt][0pt]{\rotatebox[origin=c]{90}{\textit{Field}}} &
 \includegraphics[clip,trim=8cm 0 8cm 0, width=0.2\linewidth]{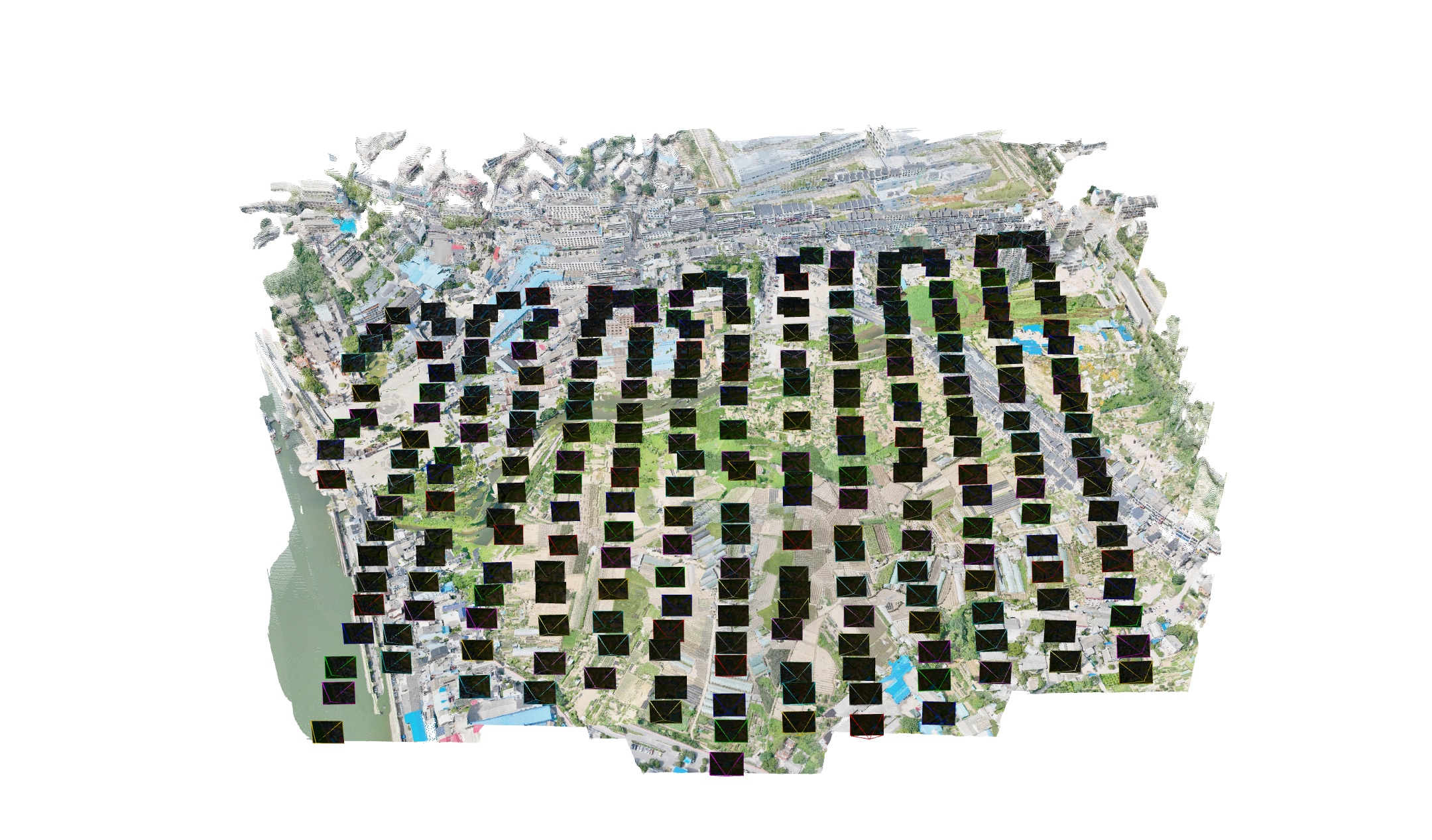} &
 \includegraphics[clip,trim=8cm 0 8cm 0, width=0.2\linewidth]{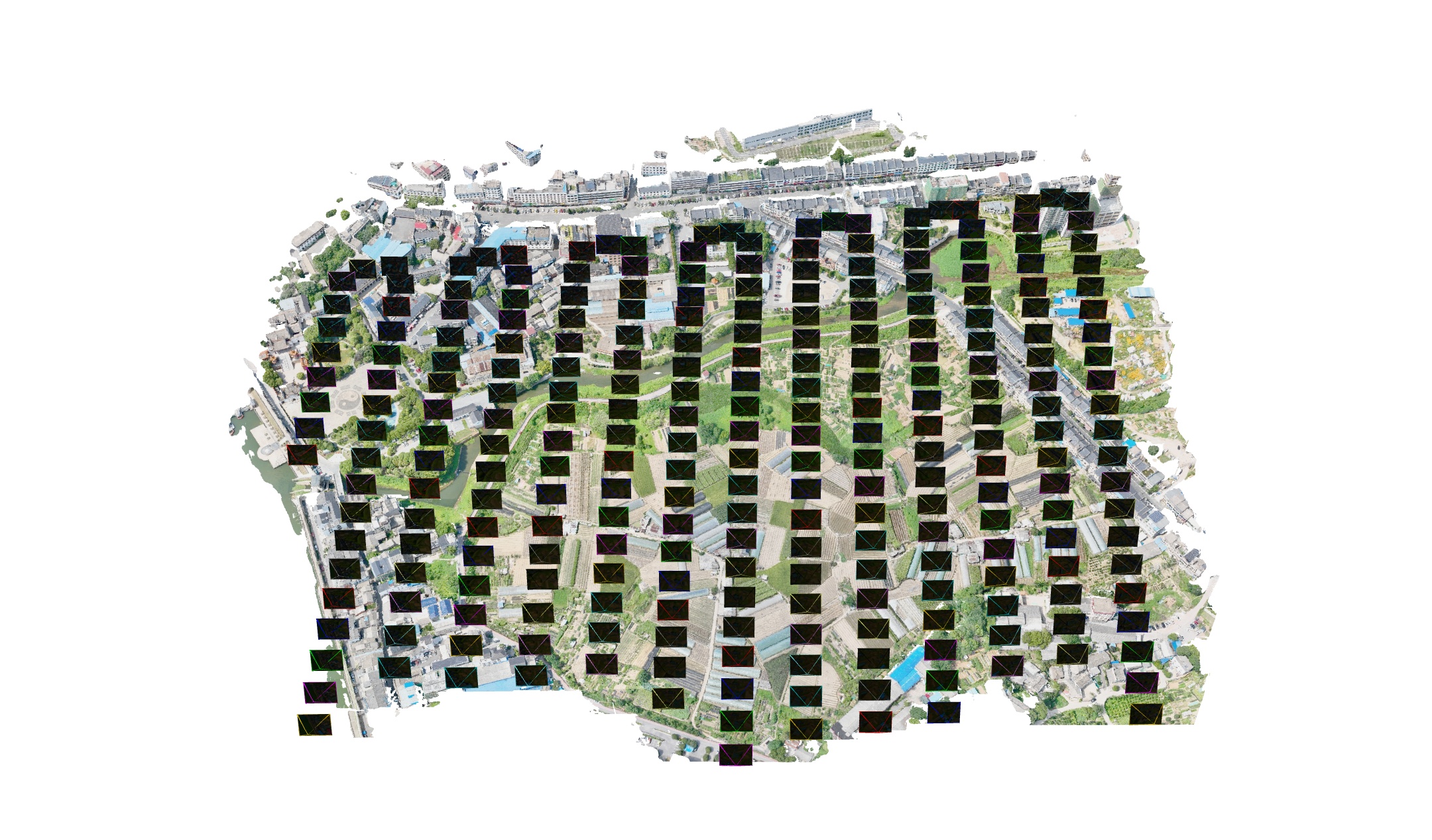} &
 \includegraphics[clip,trim=8cm 0 8cm 0, width=0.2\linewidth]{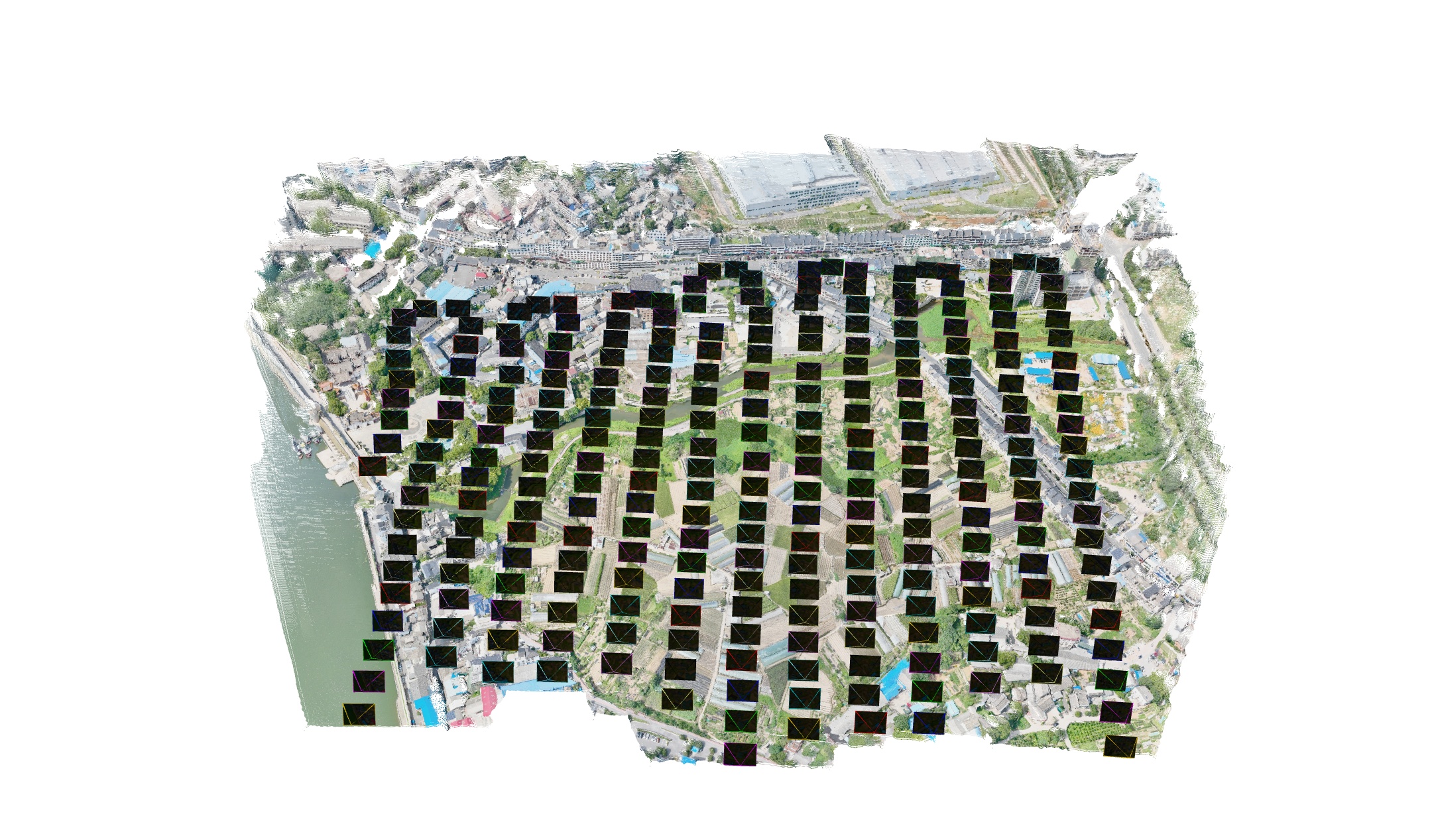} &
 \includegraphics[clip,trim=8cm 0 8cm 0, width=0.2\linewidth]{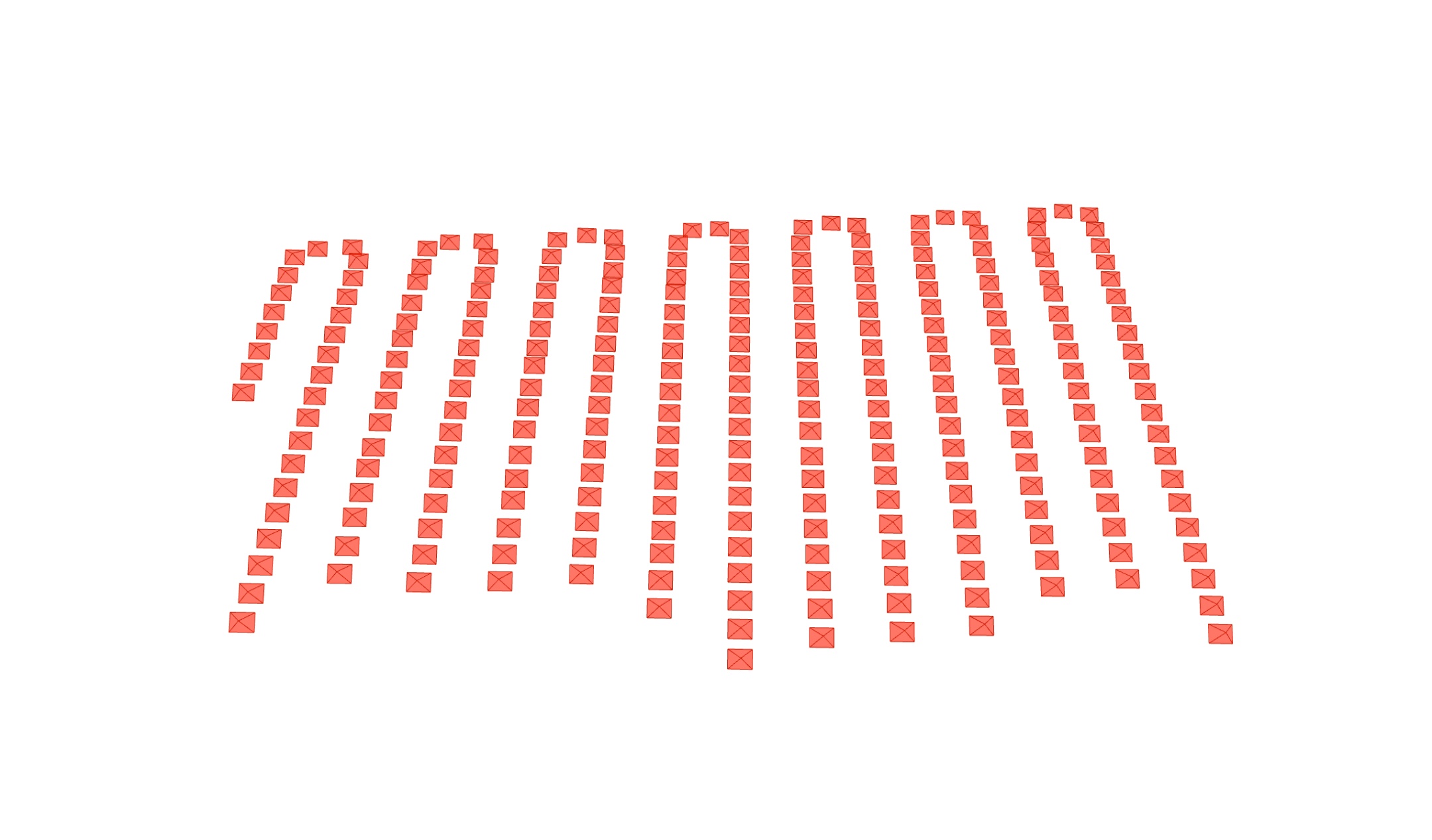} \\
 \hline
 
 \raisebox{1cm}[0pt][0pt]{\rotatebox[origin=c]{90}{\textit{Hotel}}} &
 \includegraphics[clip,trim=8cm 0 8cm 0, width=0.2\linewidth]{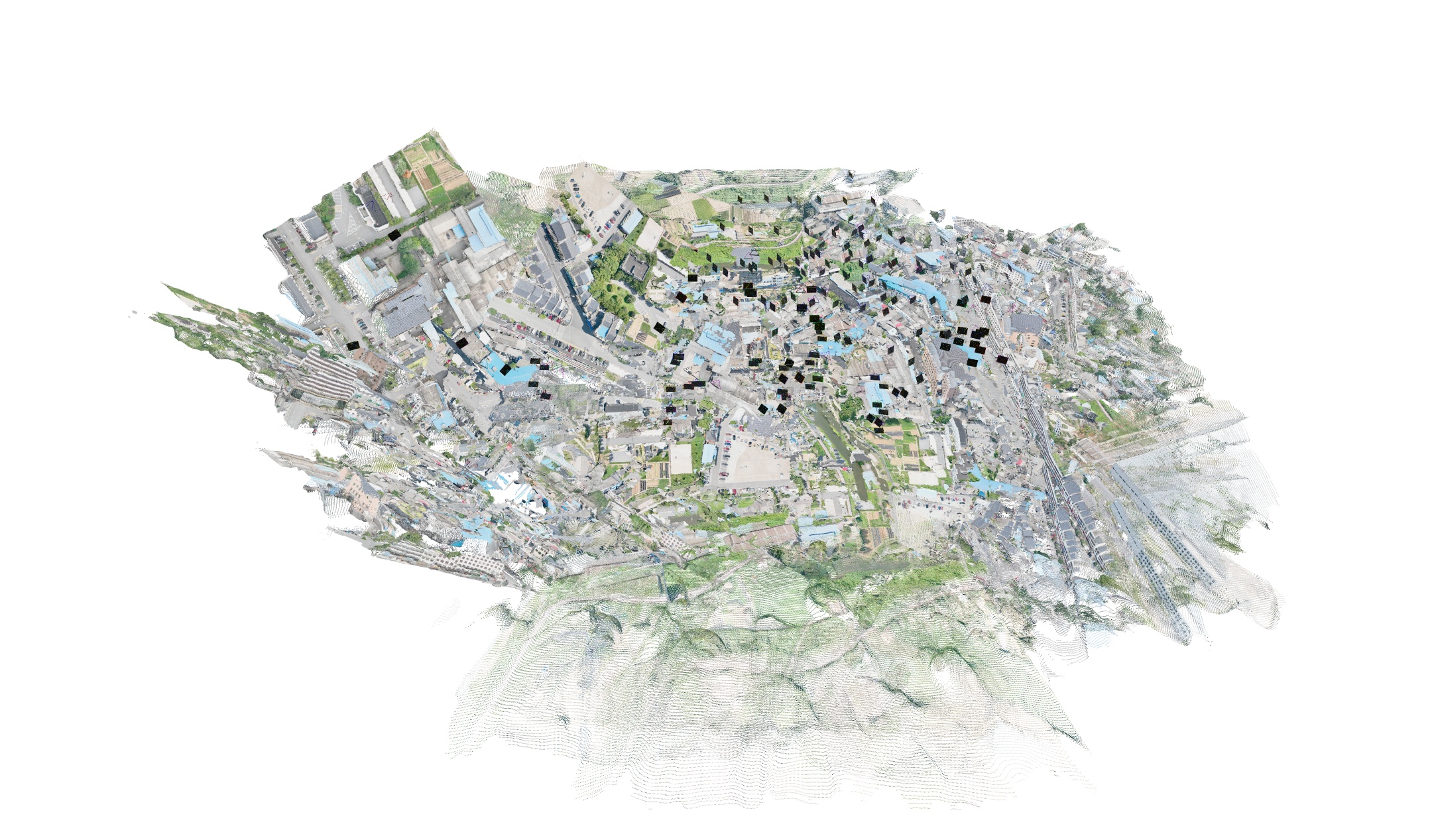} &
 \includegraphics[clip,trim=8cm 0 8cm 0, width=0.2\linewidth]{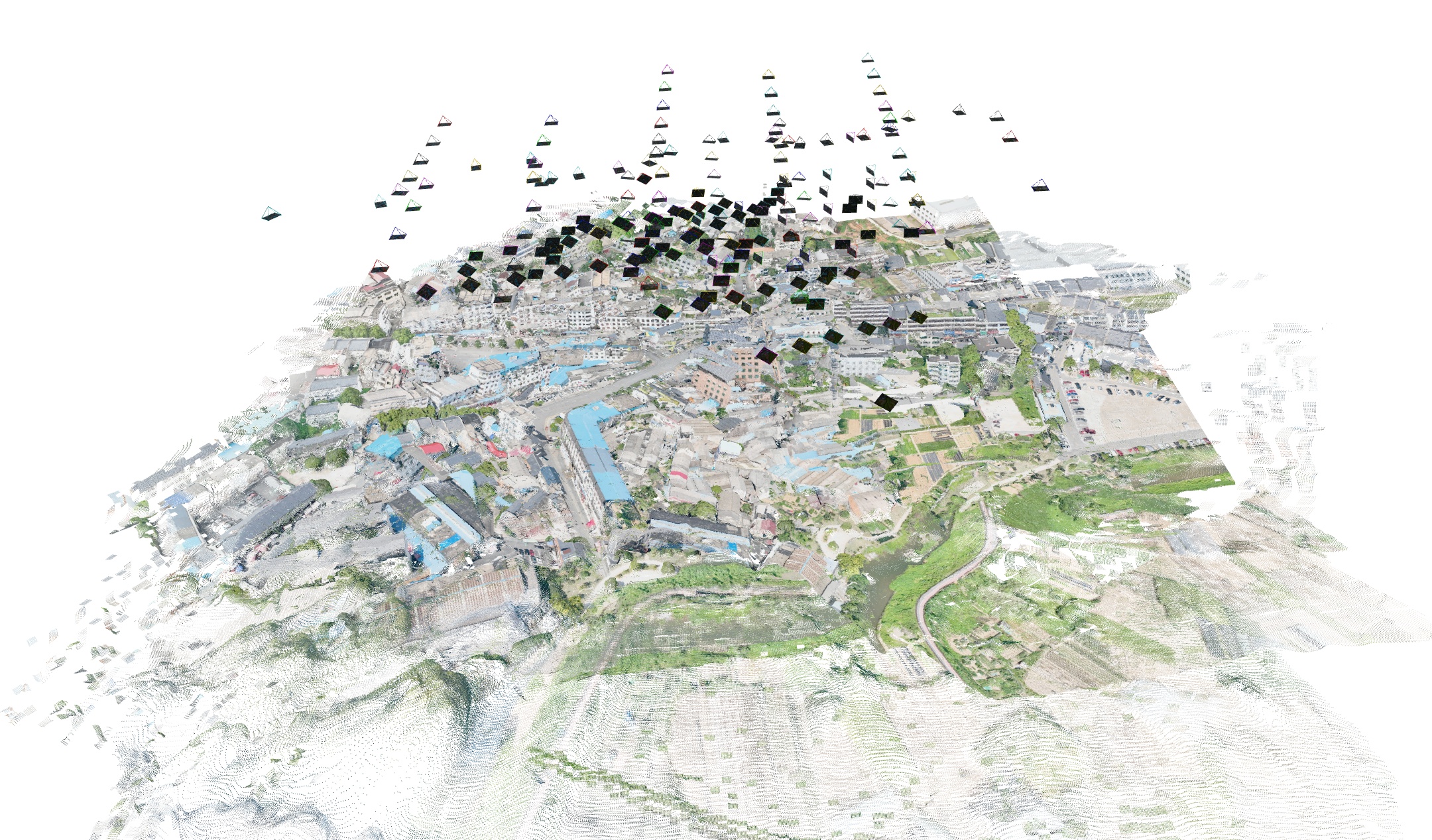} &
 \includegraphics[clip,trim=8cm 0 8cm 0, width=0.2\linewidth]{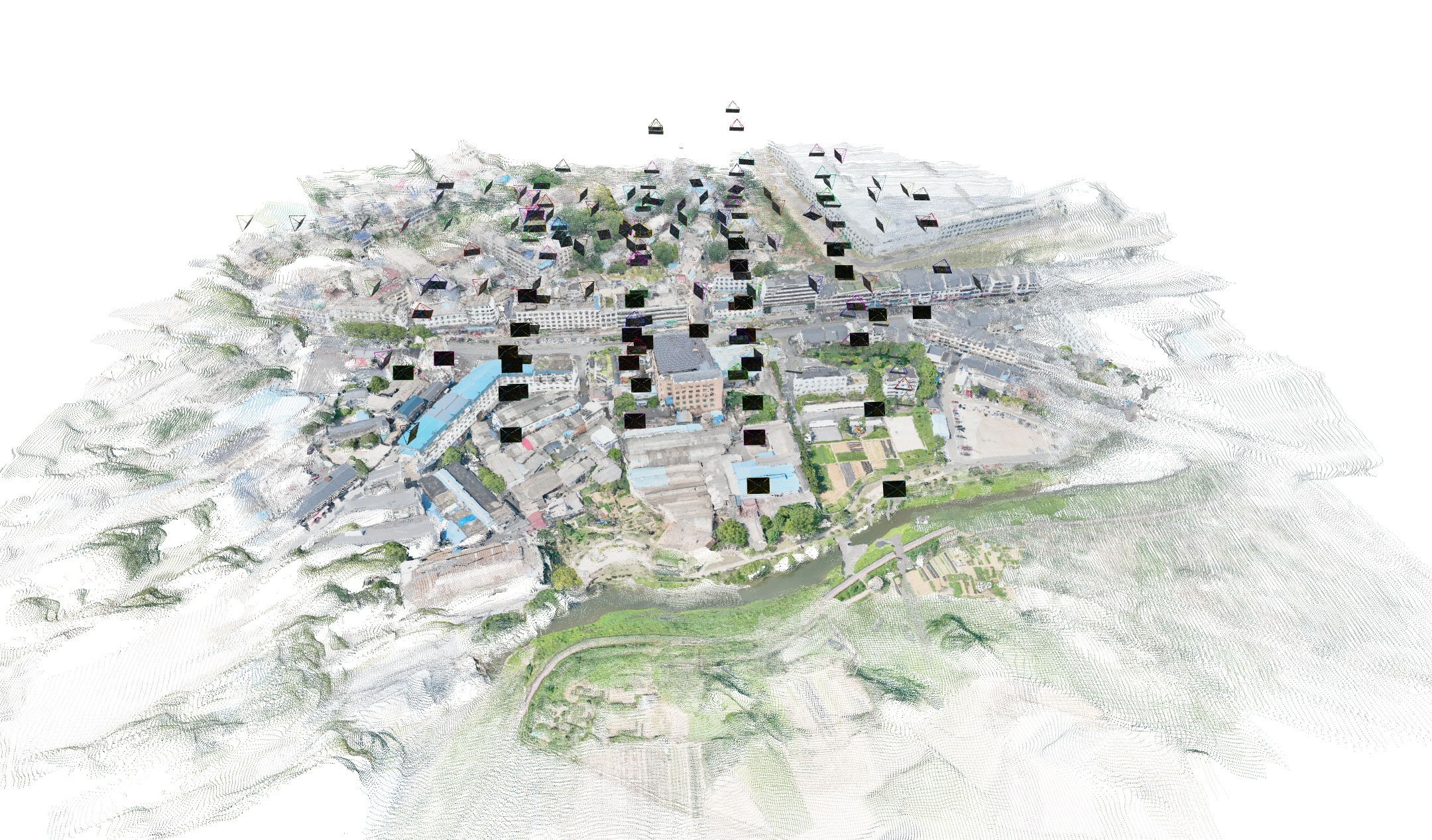} &
 \includegraphics[clip,trim=8cm 0 8cm 0, width=0.2\linewidth]{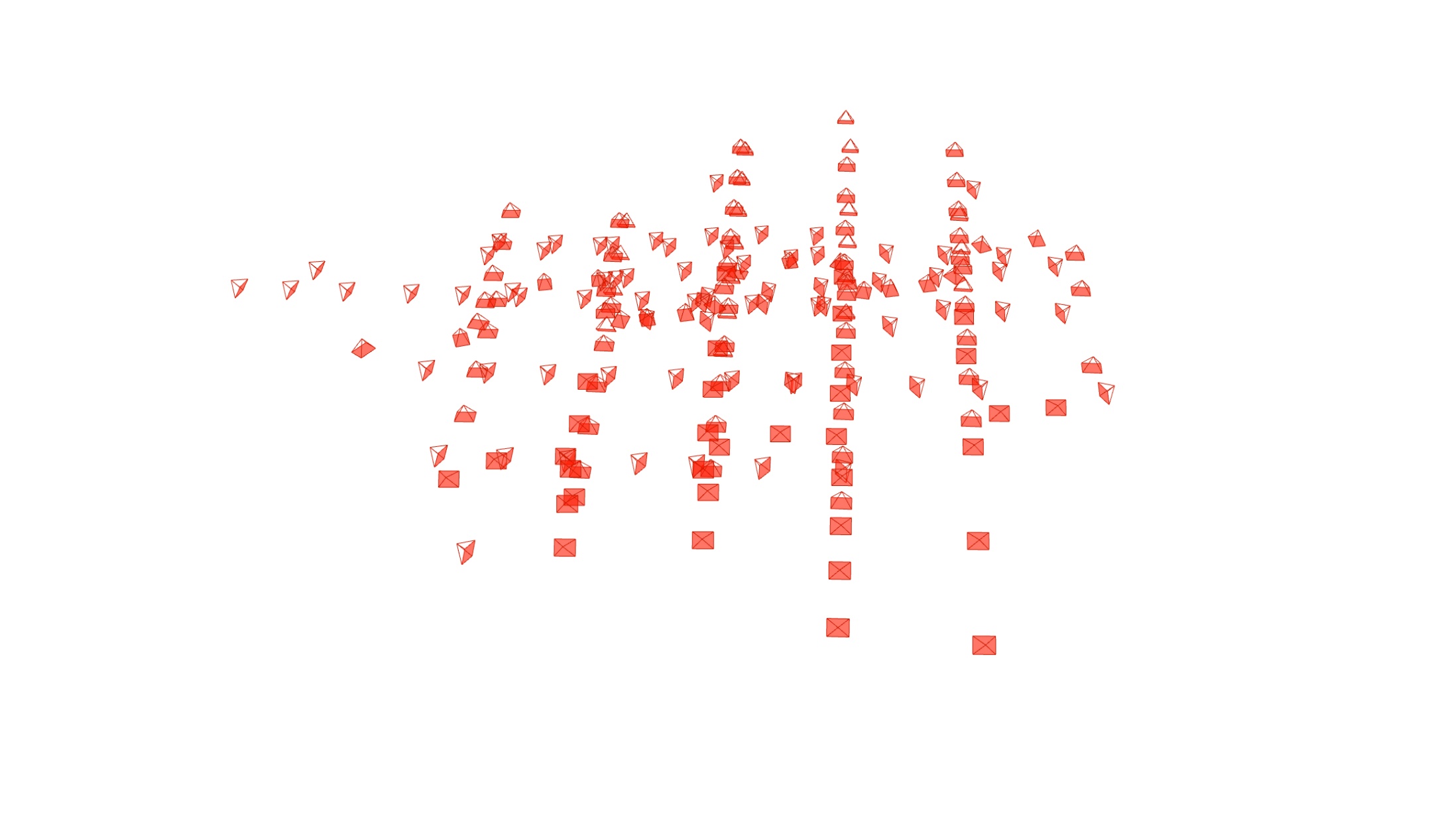} \\
 \hline
 
 \raisebox{1cm}[0pt][0pt]{\rotatebox[origin=c]{90}{\textit{Bridge}}} &
 \includegraphics[clip,trim=8cm 0 8cm 0, width=0.2\linewidth]{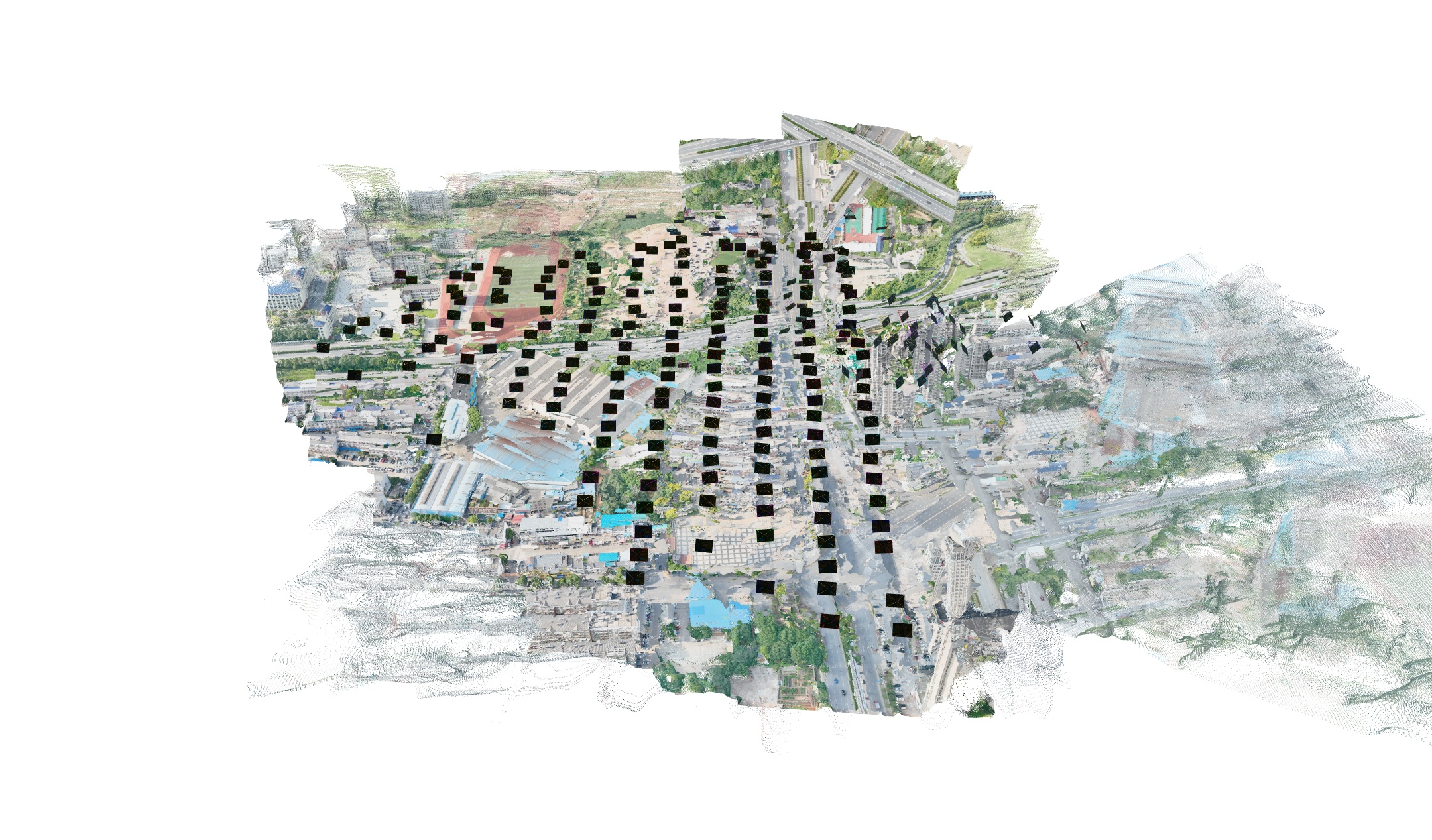} &
 \includegraphics[clip,trim=8cm 0 8cm 0, width=0.2\linewidth]{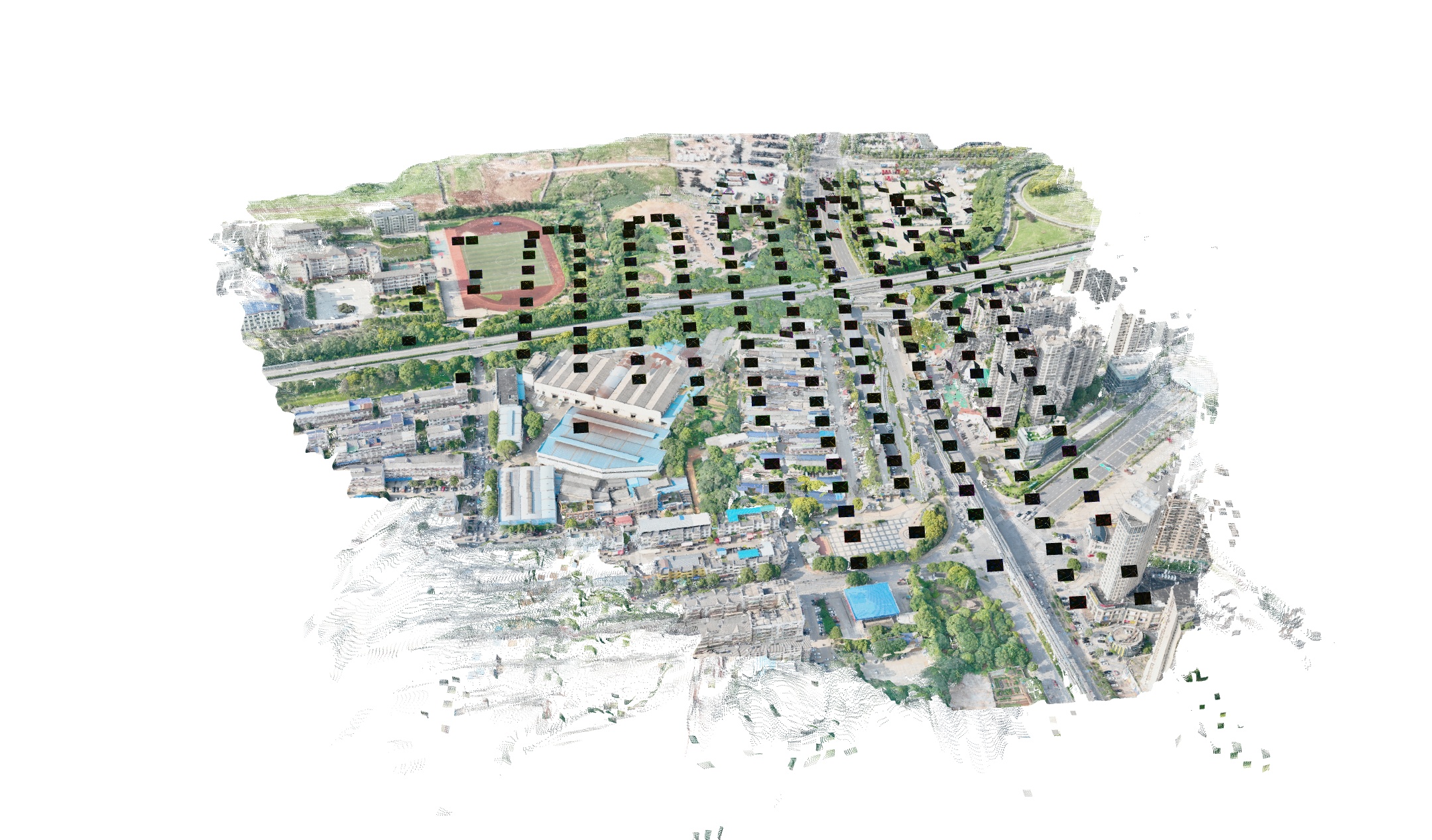} &
 \includegraphics[clip,trim=8cm 0 8cm 0, width=0.2\linewidth]{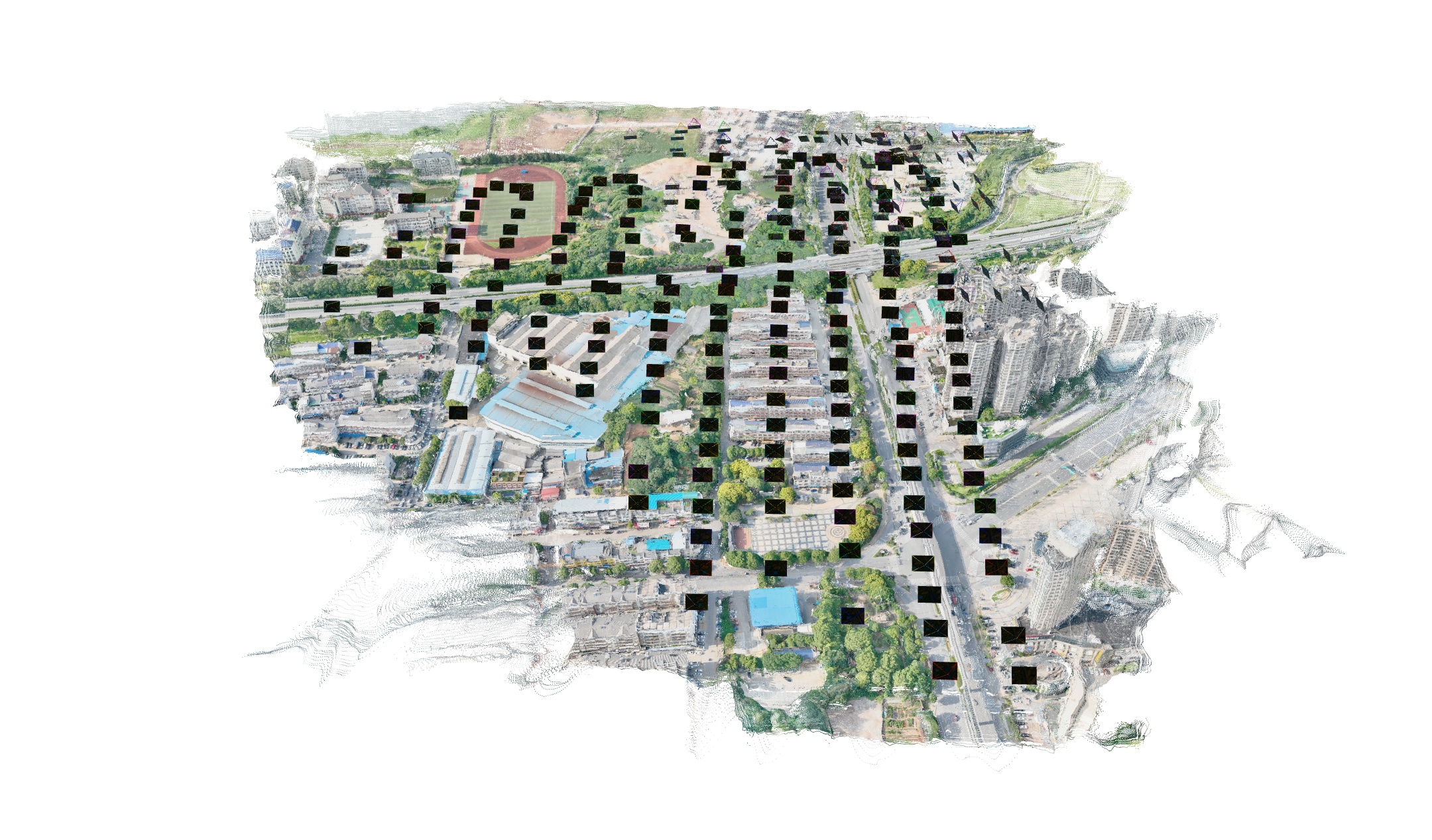} &
 \includegraphics[clip,trim=8cm 0 8cm 0, width=0.2\linewidth]{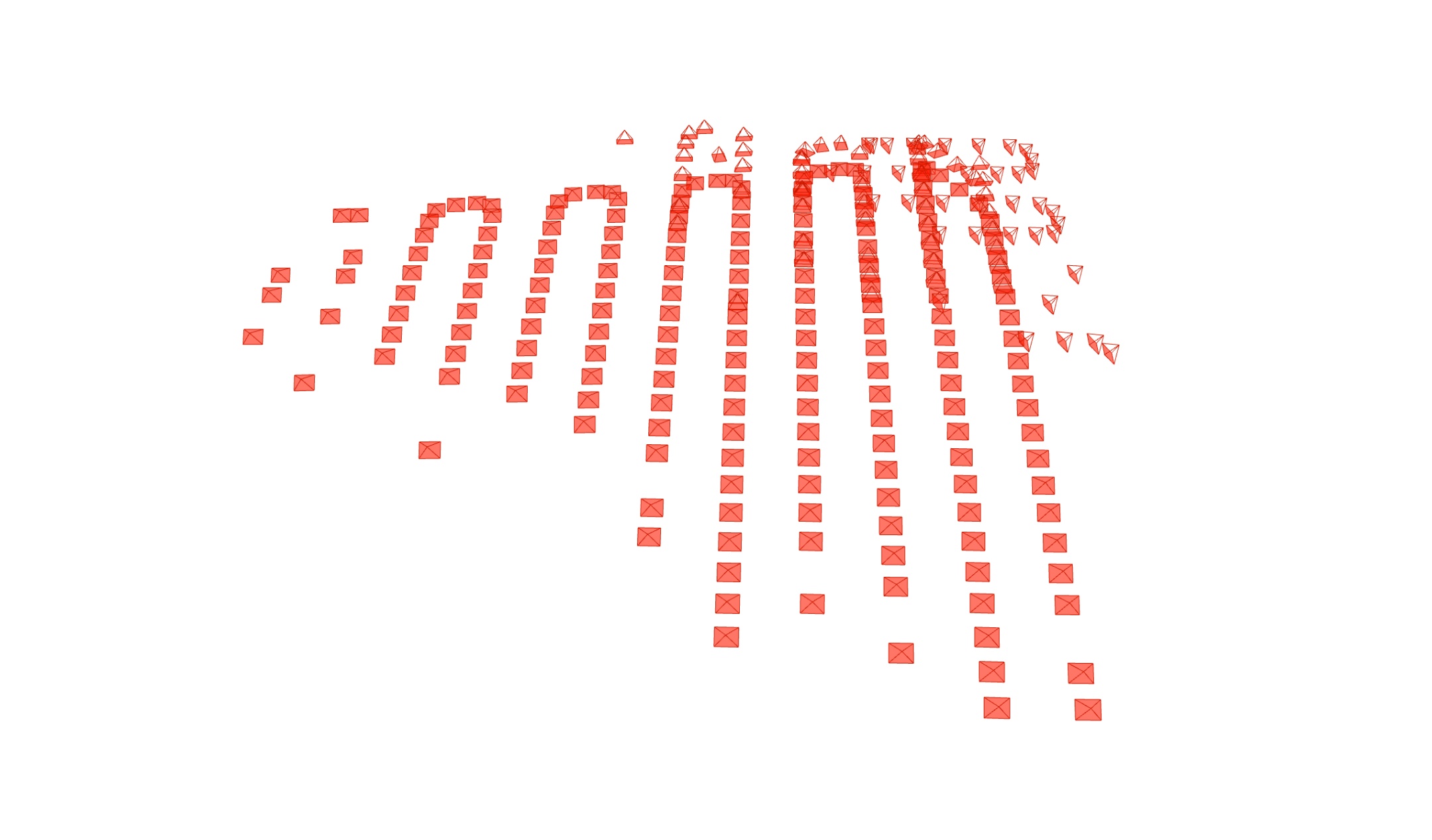} \\
 \hline

 \raisebox{1.2cm}[0pt][0pt]{\rotatebox[origin=c]{90}{\textit{Bridge (front)}}} &
 \includegraphics[clip,trim=8cm 0 8cm 0, width=0.2\linewidth]{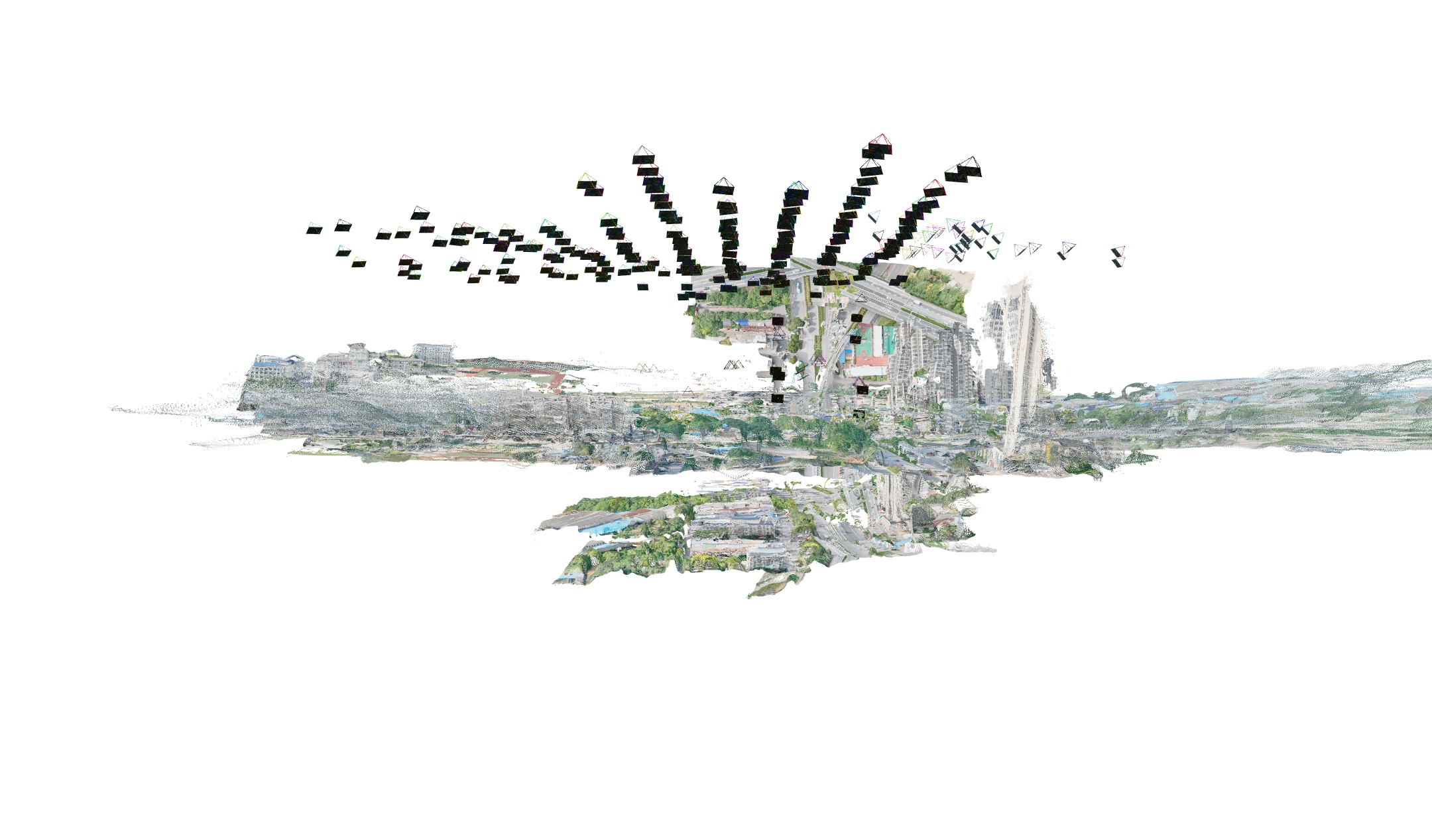} &
 \includegraphics[clip,trim=8cm 0 8cm 0, width=0.2\linewidth]{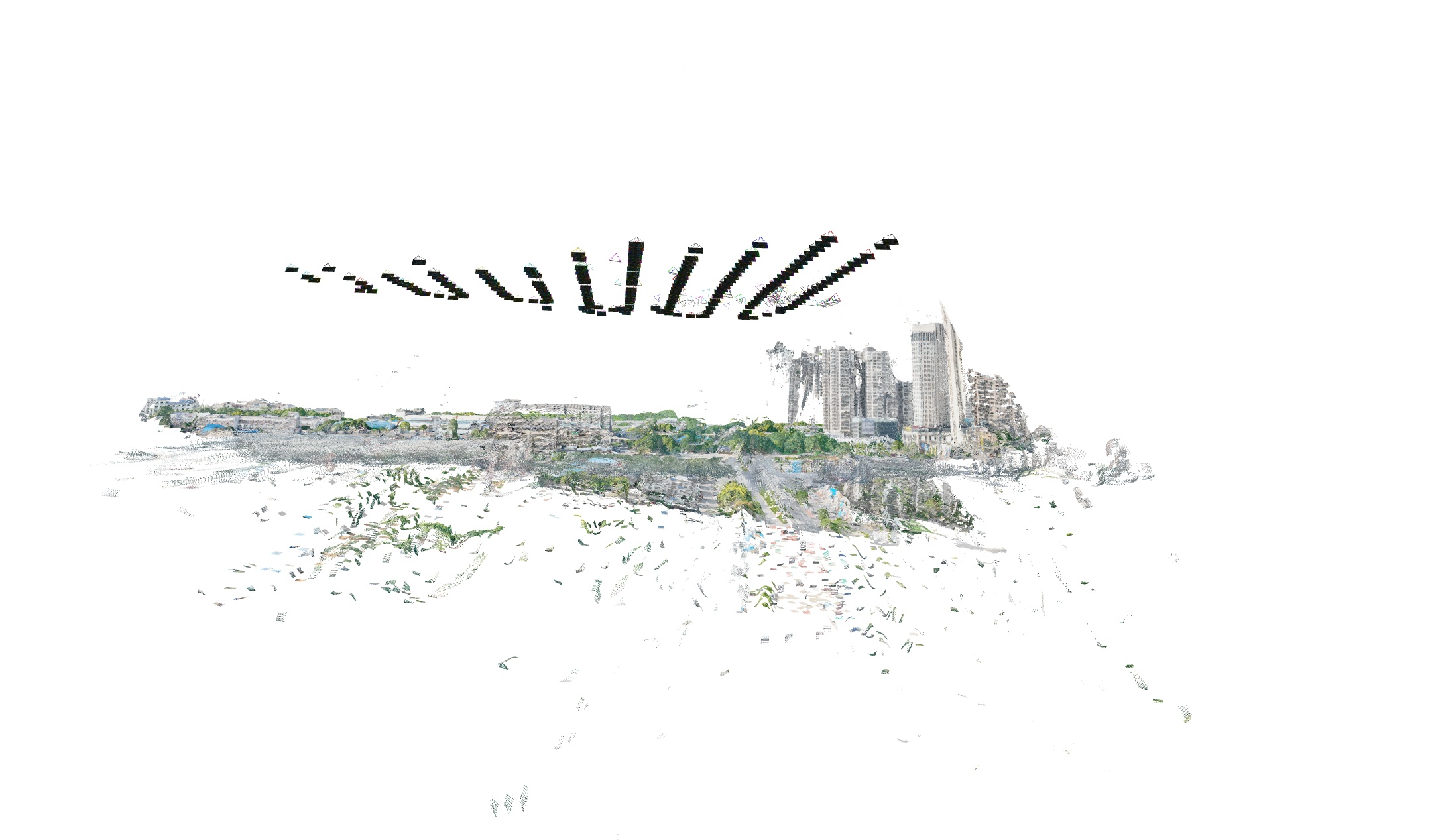} &
 \includegraphics[clip,trim=8cm 0 8cm 0, width=0.2\linewidth]{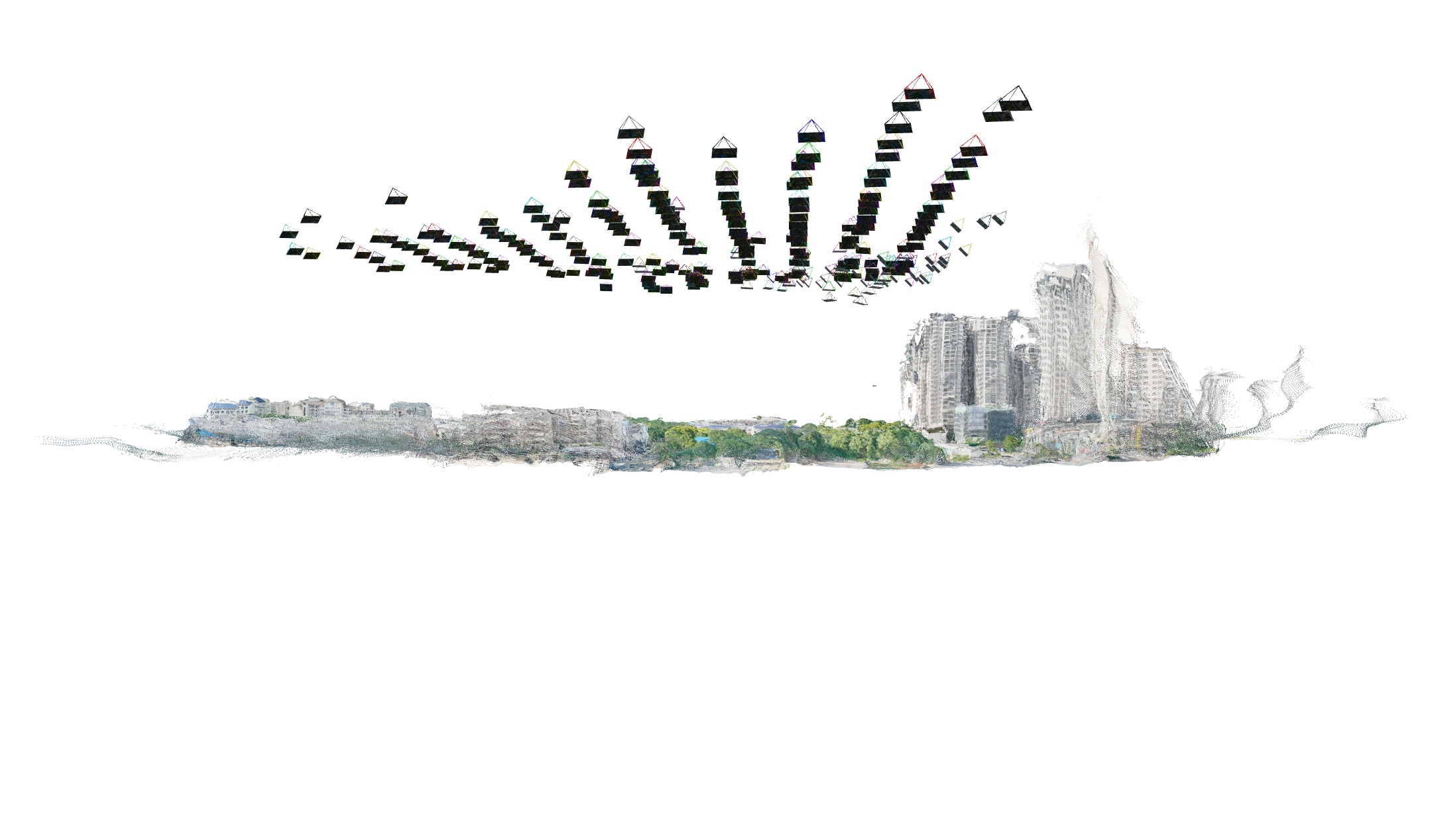} &
 \includegraphics[clip,trim=8cm 0 8cm 0, width=0.2\linewidth]{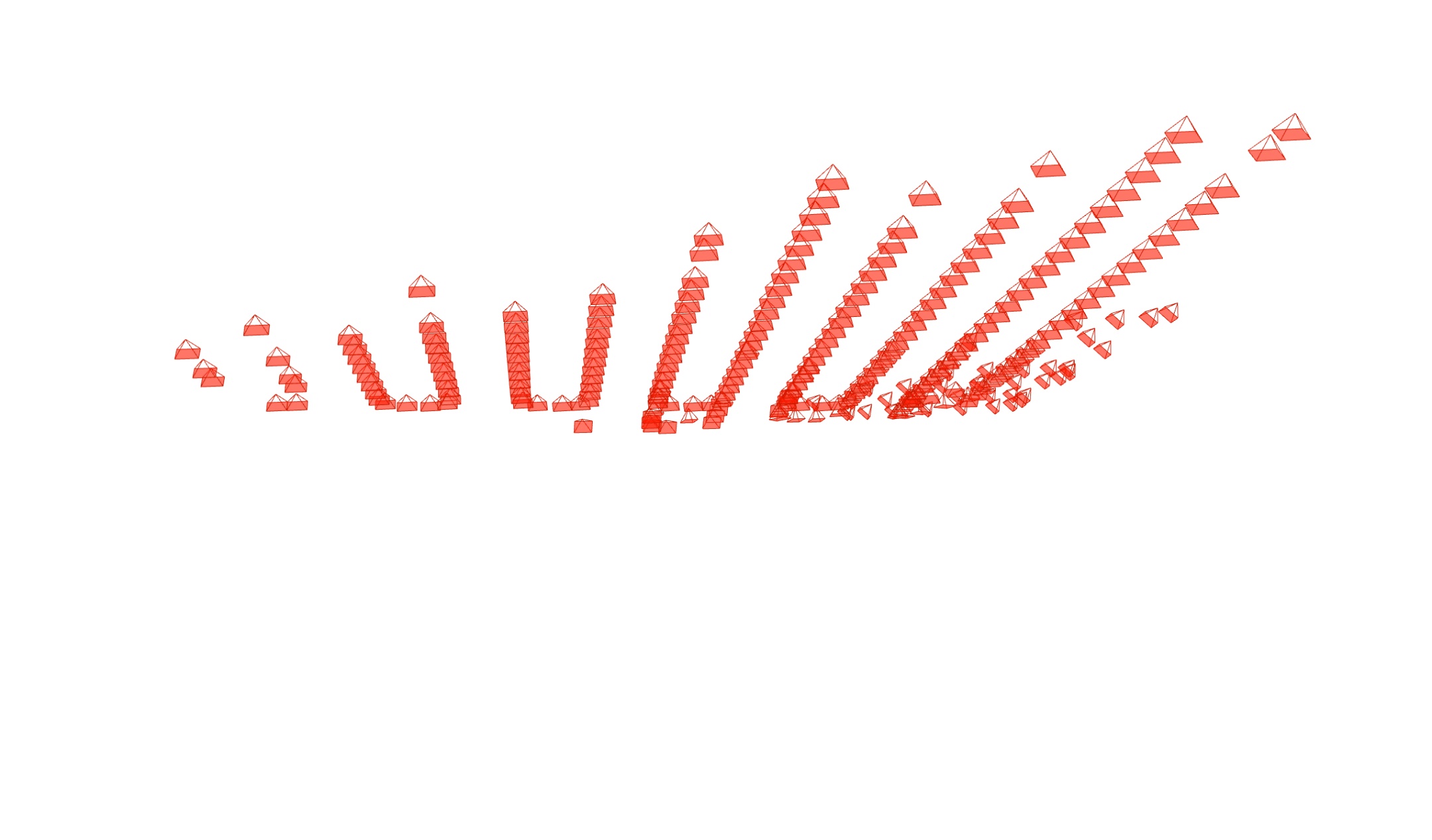} \\
 \end{tabular}
 
 \caption{Qualitative comparison of different methods across multiple scenes of CS-Drone3D dataset.}
 \label{fig:comparison}
 \end{figure*}

\section{Experiments}

This section experimentally validates the effectiveness of the proposed Regist3R. Our experiments compare Regist3R against baselines in terms of pose estimation accuracy and MVS point cloud quality, supplemented by ablation studies to verify the contribution of each module during training and inference. Furthermore, we propose an ensemble-based post-optimization strategy to enhance reconstruction accuracy, and demonstrate large-scale reconstruction of scenes with over 1,000 images within 5 minutes (Appendix Sec. 3 \& 4).

\noindent\textbf{Training Data.} Following DUSt3R, we train our model on a mix of 7 datasets: MegaDepth~\cite{li2018megadepth}, ARKitScenes~\cite{baruch2021arkitscenes}, Static Scenes 3D~\cite{schroppel2022benchmark}, Blended MVS~\cite{yao2020blendedmvs}, ScanNet++~\cite{yeshwanth2023scannet++}, CO3D-v2~\cite{reizenstein2021common}, and Waymo~\cite{sun2020scalability}, covering the indoor, outdoor, and object-centric scenes. Training details can be found in Appendix Sec. 1.

\noindent\textbf{Baselines.} The relative baselines can be categorized into post-optimization-based and inference-only. For post-optimization-based approaches, DUSt3R~\cite{wang2024dust3r} uses pointmap global alignment to align pairs. MASt3R-SfM~\cite{duisterhof2024mast3r} uses parse matching points alignment to improve scalability. This kind of approach achieves high accuracy pose estimation but takes a long time. For inference-only approaches, which are our main focus, Spann3R~\cite{wang20243d} organizes the views into a sequence and uses a memory bank to store history information. It requires the input views to be ordered. For an unordered set, the dense pairwise graph is built with DUSt3R, and a sequence is extracted according to the confidence. Fast3R~\cite{yang2025fast3r} is a recently proposed multi-view extension of DUSt3R, which receives all views as input and predicts the pointmap of each view. MV-DUSt3R~\cite{tang2024mv} is similar to Fast3R but Fast3R achieves better performance, therefore we only take Fast3R as the representative of multi-view models.

We further refer to the infer-then-align mode of Light3R-SfM~\cite{elflein2025light3r} to optimize the multi-view reconstruction efficiency and scalability of DUSt3R, which uses the common view to perform cross-pair Procrustes alignment after performing pairwise inference. We name this model DUSt3R$^\dagger$.

\noindent\textbf{Evaluation Datasets and Metrics.} We evaluate our model on three public datasets: DTU~\cite{aanaes2016large}, NRGBD~\cite{azinovic2022neural}, and 7scenes~\cite{shotton2013scene}, and a self-made dataset CS-Drone3D. DTU, NRGBD, and 7scenes are indoor or object-centric datasets with ground truth depth maps and camera poses. CS-Drone3D is an oblique aerial dataset with a large spatial span, acquired by a DJI drone at an altitude of about 140 meters. It contains three scenes: field with 258 images, hotel with 197 images, and bridge with 275 images. Some examples are shown in Appendix Sec. 2. 

Given a set of images, we follow previous works to compute the relative camera pose errors for all image pairs and measure the percentage of pairs with angular rotation/translation error below a certain threshold $\tau$, denoted as relative rotation accuracy (RRA@$\tau$) and relative translation accuracy (RTA@$\tau$). We also report mean Average Accuracy below 30 (mAA@30) defined as the area under the curve in (RRA@$\tau$, RTA@$\tau$) at a threshold $\tau$, integrated over [1, 30]. We report its accuracy score average over all data samples. We also report the runtime on a system with an NVIDIA A100 (40GB). For the datasets that provide depth maps, we further report the point cloud precision with the measurement of point cloud accuracy (Acc) and completion (Comp). We \textbf{shuffle the images} to evaluate the performance on unordered set.


\subsection{Comparative Study}

\subsubsection{Object-level reconstruction.} In Tab.~\ref{tab:benchmark-DTU}, we evaluate the object-level reconstruction on the DTU dataset. DTU is an unordered set where cameras are evenly distributed in front of and above the objects. Each scan contains 49 views and we take all views as input for reconstruction. DUSt3R is evaluated with the 224 resolution to fit the memory requirement of post-optimization. Spann3R supports 224 resolution only.
We set the tree-compression factor as 1 for Regist3R. Regist3R achieves comparable or better performance than post-optimization-based approach DUSt3R but the time cost is significantly reduced. Compared to the inference-only multi-view model Fast3R, Regist3R achieves more accurate camera pose estimation and better geometry. The time-consuming aspects of Regist3R include ASMK retrieval, MST building, and incremental reference. The time sole reference takes is comparable to Fast3R. The Spann3R is in offline mode to handle unordered image sets, therefore its performance and efficiency are limited.

\begin{table}[t]
\centering
\caption{Performance comparison on DTU dataset}
\label{tab:benchmark-DTU}
\footnotesize
\begin{tabular}{l *{4}{S[table-format=3.3]}}
\toprule
\textbf{Metric} & \textbf{DUSt3R} & \textbf{Spann3R} & \textbf{Fast3R} & \textbf{Regist3R} \\
\midrule
mAA@30     & 0.734      & 0.403      & \cellcolor{second}0.737 & \cellcolor{best}0.799 \\
RRA@5      & \cellcolor{best}0.916 & 0.272      & 0.601       & \cellcolor{second}0.679 \\
RRA@10     & \cellcolor{best}0.987 & 0.477      & 0.846       & \cellcolor{second}0.934 \\
RRA@15     & \cellcolor{best}0.994 & 0.569      & 0.925       & \cellcolor{second}0.970 \\
RTA@5      & 0.465      & 0.221      & \cellcolor{second}0.510 & \cellcolor{best}0.630 \\
RTA@10     & 0.733      & 0.428      & \cellcolor{second}0.799 & \cellcolor{best}0.871 \\
RTA@15     & 0.845      & 0.512      & \cellcolor{second}0.890 & \cellcolor{best}0.939 \\
\midrule
Acc    & \cellcolor{best}2.337 & 3.579      & 4.038       & \cellcolor{second}3.193 \\
Comp    & \cellcolor{best}2.054 & 2.527      & \cellcolor{second}2.429 & 2.441 \\
\midrule
Time (s)    & 120.671     & 114.722     & \cellcolor{best}2.050  & \cellcolor{second}7.137 \\
\bottomrule
\end{tabular}
\end{table}

\begin{table}[t]
\centering
\caption{Performance comparison on Neural RGBD dataset}
\label{tab:benchmark-nrgbd}
\footnotesize
\begin{tabular}{l *{4}{S[table-format=3.3]}}
\toprule
\textbf{Metric} & \textbf{DUSt3R} & \textbf{Spann3R} & \textbf{Fast3R} & \textbf{Regist3R} \\
\midrule
mAA@30     & 0.782      & 0.010      & \cellcolor{best}0.834  & \cellcolor{second}0.826 \\
RRA@5      & \cellcolor{best}0.987 & 0.019   & \cellcolor{second}0.854 & 0.838 \\
RRA@10     & \cellcolor{best}0.999 & 0.064   & \cellcolor{second}0.973 & \cellcolor{best}0.999 \\
RRA@15     & \cellcolor{second}0.999 & 0.112  & 0.992          & \cellcolor{best}1.000 \\
RTA@5      & 0.594      & 0.004      & \cellcolor{best}0.698  & \cellcolor{second}0.666 \\
RTA@10     & 0.795      & 0.012      & \cellcolor{second}0.859 & \cellcolor{best}0.879 \\
RTA@15     & 0.872      & 0.024      & \cellcolor{second}0.924 & \cellcolor{best}0.945 \\
\midrule
Acc    & \cellcolor{best}0.052 & 0.142   & \cellcolor{second}0.069 & 0.081 \\
Comp    & \cellcolor{best}0.033 & 0.104   & \cellcolor{second}0.039 & 0.047 \\
\midrule
Time (s)    & 67.477     & 154.577     & \cellcolor{best}2.870  & \cellcolor{second}8.576 \\
\bottomrule
\end{tabular}
\end{table}

\begin{table}[t]
\centering
\caption{Performance comparison on 7 scenes dataset}
\label{tab:benchmark-7scenes}
\footnotesize
\begin{tabular}{l *{4}{S[table-format=3.3]}}
\toprule
\textbf{Metric} & \textbf{DUSt3R} & \textbf{Spann3R} & \textbf{Fast3R} & \textbf{Regist3R} \\
\midrule
mAA@30     & 0.592      & 0.131      & \cellcolor{second}0.642 & \cellcolor{best}0.677 \\
RRA@5      & \cellcolor{best}0.611 & 0.105      & \cellcolor{second}0.606 & 0.573 \\
RRA@10     & \cellcolor{best}0.952 & 0.220      & 0.820       & \cellcolor{second}0.905 \\
RRA@15     & \cellcolor{best}0.990 & 0.315      & 0.866       & \cellcolor{second}0.966 \\
RTA@5      & 0.301      & 0.059      & \cellcolor{best}0.485  & \cellcolor{second}0.448 \\
RTA@10     & 0.551      & 0.122      & \cellcolor{second}0.666 & \cellcolor{best}0.705 \\
RTA@15     & 0.700      & 0.174      & \cellcolor{second}0.749 & \cellcolor{best}0.807 \\
\midrule
Acc    & \cellcolor{best}0.033 & 0.064      & 0.051       & \cellcolor{second}0.035 \\
Comp    & \cellcolor{best}0.034 & 0.110      & 0.057       & \cellcolor{second}0.040 \\
\midrule
Time (s)    & 39.189     & 85.497      & \cellcolor{best}1.789  & \cellcolor{second}6.724 \\
\bottomrule
\end{tabular}
\end{table}

\begin{table}[t]
\centering
\caption{Performance Comparison on CS-Drone3D Dataset}
\label{tab:benchmark-CS-Drone3D}
\sisetup{
  separate-uncertainty,
  round-mode = places,
  round-precision = 3  
}
\footnotesize
\begin{tabular}{l|l|*{3}{S|}S}
\toprule
\multicolumn{1}{c|}{\textbf{Scene}} & \multicolumn{1}{c|}{\textbf{Metric}} & \multicolumn{1}{c}{\textbf{DUSt3R$^\dagger$}} & \multicolumn{1}{c}{\textbf{Fast3R}} & \multicolumn{1}{c}{\textbf{MASt3R-SfM}} & \multicolumn{1}{c}{\textbf{Regist3R}} \\
\midrule
\multirow{8}{*}{\textbf{Field}} 
& mAA@30 & 0.276 & 0.049 & \cellcolor{best}0.915 & \cellcolor{second}0.876 \\
& RRA@5 & \cellcolor{second}0.810 & 0.033 & \cellcolor{best}0.963 & 0.782 \\
& RRA@10 & \cellcolor{second}0.956 & 0.064 & \cellcolor{best}1.000 & \cellcolor{best}1.000 \\
& RRA@15 & \cellcolor{second}0.983 & 0.108 & \cellcolor{best}1.000 & \cellcolor{best}1.000 \\
& RTA@5 & 0.196 & 0.021 & \cellcolor{second}0.882 & \cellcolor{best}0.896 \\
& RTA@10 & 0.275 & 0.048 & \cellcolor{second}0.992 & \cellcolor{best}0.992 \\
& RTA@15 & 0.313 & 0.080 & \cellcolor{best}0.998 & \cellcolor{second}0.998 \\
& Time(s) & \cellcolor{best}25.834 & 27.913 & 536.626 & \cellcolor{second}25.882 \\
\midrule

\multirow{8}{*}{\textbf{Hotel}}
& mAA@30 & 0.628 & 0.175 & \cellcolor{second}0.691 & \cellcolor{best}0.799 \\
& RRA@5 & 0.661 & 0.067 & \cellcolor{second}0.878 & \cellcolor{best}0.896 \\
& RRA@10 & 0.710 & 0.177 & \cellcolor{best}1.000 & \cellcolor{second}0.961 \\
& RRA@15 & 0.731 & 0.281 & \cellcolor{best}1.000 & \cellcolor{second}0.970 \\
& RTA@5 & \cellcolor{second}0.557 & 0.067 & 0.417 & \cellcolor{best}0.612 \\
& RTA@10 & \cellcolor{second}0.694 & 0.180 & 0.655 & \cellcolor{best}0.848 \\
& RTA@15 & 0.735 & 0.275 & \cellcolor{second}0.786 & \cellcolor{best}0.913 \\
& Time(s) & \cellcolor{second}19.788 & \cellcolor{best}17.141 & 447.721 & 20.667 \\
\midrule

\multirow{8}{*}{\textbf{Bridge}}
& mAA@30 & 0.530 & 0.262 & \cellcolor{second}0.789 & \cellcolor{best}0.842 \\
& RRA@5 & 0.601 & 0.061 & \cellcolor{second}0.733 & \cellcolor{best}0.831 \\
& RRA@10 & 0.770 & 0.225 & \cellcolor{best}0.996 & \cellcolor{second}0.978 \\
& RRA@15 & 0.777 & 0.371 & \cellcolor{best}1.000 & \cellcolor{second}0.978 \\
& RTA@5 & 0.479 & 0.104 & \cellcolor{second}0.661 & \cellcolor{best}0.788 \\
& RTA@10 & 0.601 & 0.294 & \cellcolor{second}0.817 & \cellcolor{best}0.927 \\
& RTA@15 & 0.643 & 0.446 & \cellcolor{second}0.893 & \cellcolor{best}0.955 \\
& Time(s) & \cellcolor{best}26.847 & 31.579 & 708.029 & \cellcolor{second}27.578 \\
\midrule

\multirow{8}{*}{\textbf{Average}}
& mAA@30 & 0.478 & 0.162 & \cellcolor{second}0.798 & \cellcolor{best}0.839 \\
& RRA@5 & 0.690 & 0.054 & \cellcolor{best}0.858 & \cellcolor{second}0.836 \\
& RRA@10 & 0.813 & 0.155 & \cellcolor{best}0.999 & \cellcolor{second}0.980 \\
& RRA@15 & 0.830 & 0.254 & \cellcolor{best}1.000 & \cellcolor{second}0.983 \\
& RTA@5 & 0.410 & 0.064 & \cellcolor{second}0.653 & \cellcolor{best}0.765 \\
& RTA@10 & 0.524 & 0.174 & \cellcolor{second}0.821 & \cellcolor{best}0.923 \\
& RTA@15 & 0.563 & 0.267 & \cellcolor{second}0.892 & \cellcolor{best}0.956 \\
& Time(s) & \cellcolor{best}24.156 & 25.544 & 564.125 & \cellcolor{second}24.709 \\
\bottomrule
\end{tabular}
\vspace{-0.4cm}
\end{table}

\subsubsection{Indroor scenes reconstruction.} In Tab.~\ref{tab:benchmark-nrgbd} and Tab.~\ref{tab:benchmark-7scenes}, we evaluate the performance on indoor scenes reconstruction. Both datasets are extracted from sequential video but are shuffled to meet unorder setting. We extract keyframes at the interval of 30 frames for DUSt3R to avoid memory issues, while 20 for other models. DUSt3R and Spann3R are evaluated with 224 resolution. We set the tree-compression factor as 1 for Regist3R. From the tables, we can see that Regist3R achieves comparable or better performance than DUSt3R, and is comparable to Fast3R. Fast3R is mainly trained on indoor datasets, meanwhile, the global field of view helps it perceive indoor structures, thereby achieving a better performance. In comparison, Regist3R only relies on the binocular field of view and achieves the same or even higher pose accuracy, thus proving its effectiveness.

\subsubsection{Outdoor aerial reconstruction.} In Tab.~\ref{tab:benchmark-CS-Drone3D} we evaluate the performance of outdoor aerial reconstruction. The CS-Drone3D dataset is an unordered set with a large spatial span. As the scene contains hundreds of views, DUSt3R and offline-Spann3R meet memory issues. Therefore, we compare Fast3R and MASt3R-SfM, as well as infer-then-align version DUSt3R$^\dagger$. MASt3R-SfM is post-optimized based on sparse matching points and thus scales to larger image sets, however, still suffers low efficiency. Regist3R achieves comparable performance to MASt3R-SfM but with much higher efficiency. Fast3R extends its architecture to fit large image sets, but the performance is limited. 
DUSt3R$^\dagger$ achieves efficient reconstruction but suffers from misalignment. The inference pipeline of DUSt3R$^\dagger$ is the same as Regist3R, the only difference is DUSt3R$^\dagger$ adopts infer-then-align but Regist3R adopts direct registeration. The comparison between the two shows that the alignment error affects the accuracy, while Regist3R's direct registration eliminates this error.

The qualitative comparison on the CS-Drone3D dataset is illustrated in Fig.~\ref{fig:comparison}. In relatively homogeneous \textit{field} environments, the MASt3R-SfM framework demonstrates superior feature matching accuracy, enabling robust recovery of camera poses. However, its performance degrades significantly in architecturally dense scenarios such as \textit{hotel} and \textit{bridge} structures, where diminished feature matching precision leads to reconstruction failures. In contrast, the Regist3R approach circumvents feature matching limitations by directly estimating the target pointmap relative to reference pointmaps. Comparative baselines including Spann3R and Fast3R exhibited complete reconstruction failures across evaluated scenarios and are consequently excluded from presentation.

\begin{figure}
 \centering
 \includegraphics[width=0.8\linewidth]{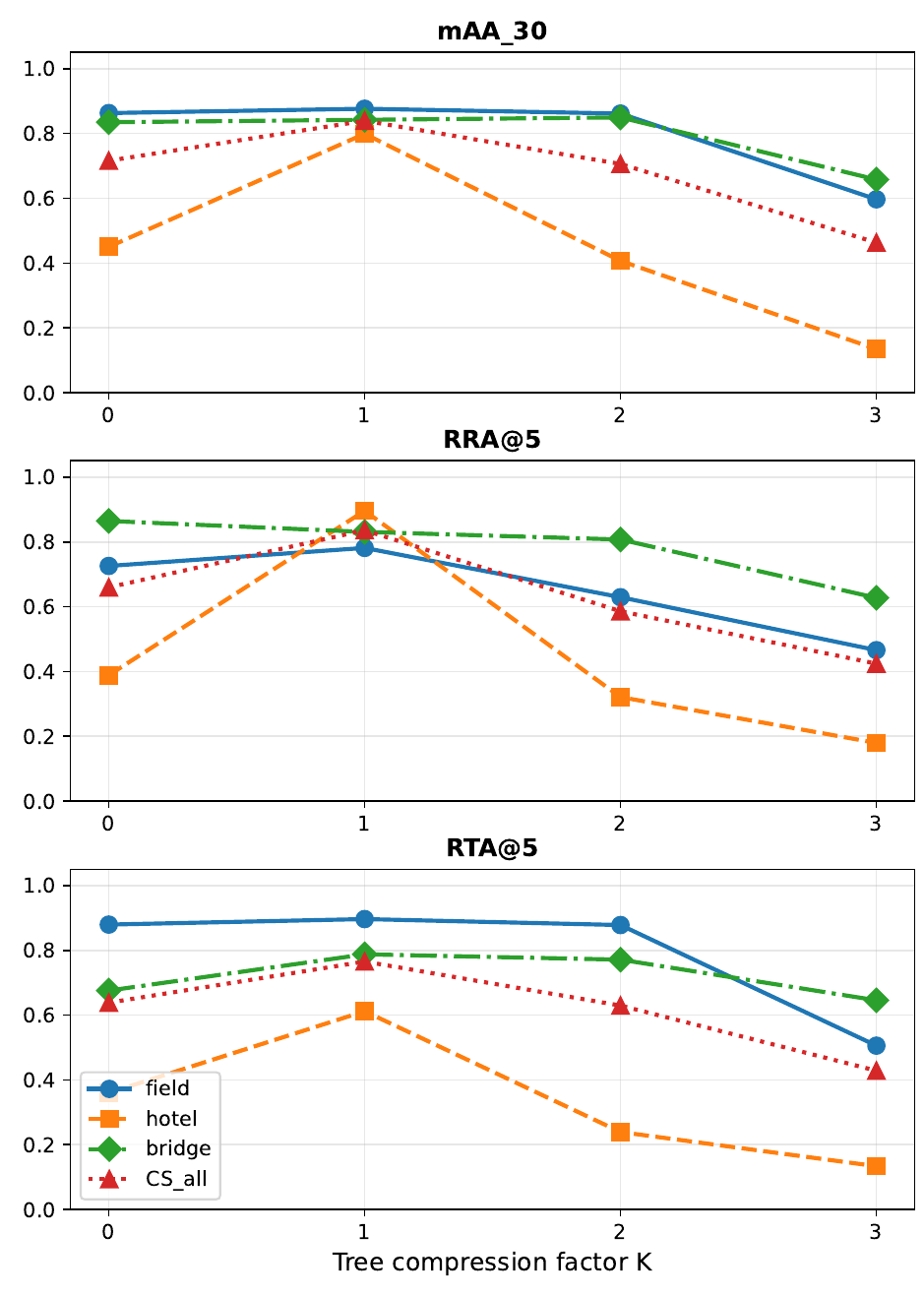}
 \caption{Performance of Regist3R with different tree compression factors. The depth of the tree is compressed to $1/2^K$ of the original depth.}
 \label{fig:tree compression factor}
 \vspace{-0.5cm}
\end{figure}

\subsection{Ablation Study}

We conduct an ablation study on the CS-Drone3D dataset to evaluate the effectiveness of confidence-aware autoregressive training and our tree-building strategy.

\begin{table}[t]
\centering
\caption{Ablation Study on confidence-aware autoregressive training. The \textit{w/o AR} is evaluated with the same model but the confidence values are filled with all ones.}
\label{tab:autoregression}
\begin{tabular}{lcccccc}
\toprule
\textbf{scene} & \multicolumn{2}{c}{\textbf{mAA@30}} & \multicolumn{2}{c}{\textbf{RTA@5}} & \multicolumn{2}{c}{\textbf{RRA@5}} \\
\cmidrule(lr){2-3} \cmidrule(lr){4-5} \cmidrule(lr){6-7}
 & w/ AR & w/o AR & w/ AR & w/o AR & w/ AR & w/o AR \\
\midrule
field    & \textbf{0.876} & 0.860 & \textbf{0.896} & 0.846 & \textbf{0.781} & 0.685 \\
hotel    & \textbf{0.799} & 0.744 & \textbf{0.612} & 0.521 & \textbf{0.895} & 0.758 \\
bridge   & \textbf{0.842} & 0.830 & \textbf{0.787} & 0.761 & \textbf{0.830} & 0.770 \\
\midrule
Avg.   & \textbf{0.839} & 0.811 & \textbf{0.765} & 0.709 & \textbf{0.836} & 0.737 \\
\bottomrule
\end{tabular}
\end{table}

\begin{table}[t]
\centering
\caption{Ablation study on tree type}
\label{tab:mst_vs_spt}
\begin{tabular}{lcccccc}
\toprule
\textbf{scene} & \multicolumn{2}{c}{\textbf{mAA@30}} & \multicolumn{2}{c}{\textbf{RTA@5}} & \multicolumn{2}{c}{\textbf{RRA@5}} \\
\cmidrule(lr){2-3} \cmidrule(lr){4-5} \cmidrule(lr){6-7}
& MST & SPT & MST & SPT & MST & SPT \\
\midrule
field    & \textbf{0.876} & 0.172 & \textbf{0.896} & 0.090 & 0.781 & \textbf{0.922} \\
hotel   & \textbf{0.799} & 0.415 & \textbf{0.612} & 0.355 & \textbf{0.896} & 0.395 \\
bridge   & \textbf{0.842} & 0.380 & \textbf{0.788} & 0.312 & \textbf{0.831} & 0.634 \\
\midrule
Avg.  & \textbf{0.839} & 0.322 & \textbf{0.765} & 0.252 & \textbf{0.836} & 0.650 \\
\bottomrule
\end{tabular}
\end{table}

\subsubsection{Confidence-aware autoregressive training}

We evaluate the performance of Regist3R with and without confidence-aware autoregressive training. Fot a fair comparison, we reuse the same model parameters instead of retraining, but fill the confidence values with all ones in the non-autoregressive model.
The results are shown in Tab.~\ref{tab:autoregression}. The confidence-aware autoregressive training consistently improves the performance of Regist3R, especially on the RRA@5 metric. The confidence-aware autoregressive training helps the model to resist the noise and alleviate the drifting problem.

\subsubsection{Tree building strategy}

Previous work~\cite{elflein2025light3r} proposes to use the shortest path tree (SPT) instead of the commonly used minimum spanning tree (MST). However, our experiments show that this method fails to reconstruct scenes due to large spacing between pairs in scenarios with large spatial spans. The results are shown in Tab.~\ref{tab:mst_vs_spt}, where the tree compression factor is set to 1 for MST. The MST achieves better performance than SPT on all metrics. The MST is more robust to the large spacing between pairs and is more suitable for outdoor aerial reconstruction.

We then evaluate the performance of Regist3R with different tree compression factors. The results are shown in Fig.~\ref{fig:tree compression factor}. The tree compression factor K determines how many times the depth of the tree is halved, that is, the depth of the tree is $1/2^K$ of the original depth.
The performance of Regist3R is stable with different tree compression factors $K$, and the $K=1$ setting achieves the best performance. The $K=2$ setting achieves comparable performance to the original tree, while the performance drops with larger $K$. The tree compression factor of 1 is recommended for most scenes.

According to the experiments, the structure of the tree is crucial for the performance of Regist3R. The included strategies, such as MST, SPT, and tree compression, are all heuristic methods. An optimization or search-based method needs to be proposed to obtain the optimal reconstruction sequence.

\section{Conclusion}

We propose a stereo foundation model named Regist3R to address the incremental registration problem in multi-view reconstruction. Regist3R is an inference-only model that directly predicts the pointmap of the target view in the world coordinate system based on a known reference view. The experiments show its effectiveness and scalability.

There remains substantial work to be done towards modern incremental reconstruction. For example, the heuristic tree-building strategy should be replaced by a more optimal method. Modern bundle adjustment should be developed to address the drifting that inherently exists in incremental registration. The combination of the global multi-view model and incremental stereo model should be explored to achieve truly scalable and accurate multi-view reconstruction. We hope our work can inspire more research in this direction.

\begin{acks}
This work is supported by the National Nature Science Foundation of China (NSFC) under Grant No. 61902415, and the Open Fund of Science and Technology on Parallel and Distributed Processing Laboratory (PDL) under Grant No. WDZC20235250106.
\end{acks}

\bibliographystyle{ACM-Reference-Format}
\bibliography{sample-base}

\appendix

\section{Training Details} 
In our experiment, we use two ViT-L~\cite{ranftl2021vision} transformers for the reference encoder and target encoder, and two ViT-B transformers for the reference decoder and target decoder, as well as a DPT output head for target view pointmap regression. As we freeze the target encoder, the modules actually trained are one ViT-L encoder, two ViT-B decoders, and a DPT head.

Analogous to DUSt3R, a three-stage training is conducted. The parameters are initialized from the released DUSt3R model. We train our model on 8 A100 (40G) GPUs.
\begin{itemize}
  \item For the first stage, we train the model on 224 resolution with a linear head. The model is trained with a batch size of 32 per GPU for 90 epochs, 700k pairs per epoch. It takes about 3 days.
  \item For the second stage, we train the model on 512 resolution with a linear head. The model is trained with a batch size of 16 per GPU for 90 epochs, 70k pairs per epoch. It takes about 1 day.
  \item For the third stage, we train the model on 512 resolution with a DPT head and autoregressive training enabled, The model is trained with batch size of 8 per GPU for 90 epochs, 17.5k group of views per epoch. The chain length is 5. It takes about 2 days.
\end{itemize}
The total training takes about 6 days. Checkpointing is enbled to enlarge batch size.

The coefficient $\alpha$ of confidence aware regression is set to 0.5, which is 2.5 times larger than DUSt3R. This is because the pointmap is zero-centered rather than depth-normalized so the average scale is about 2.5 times larger. Other training details are kept the same as DUSt3R.

\begin{figure*}[ht]
    \centering
    \begin{subfigure}[b]{0.195\textwidth}
        \includegraphics[width=\textwidth]{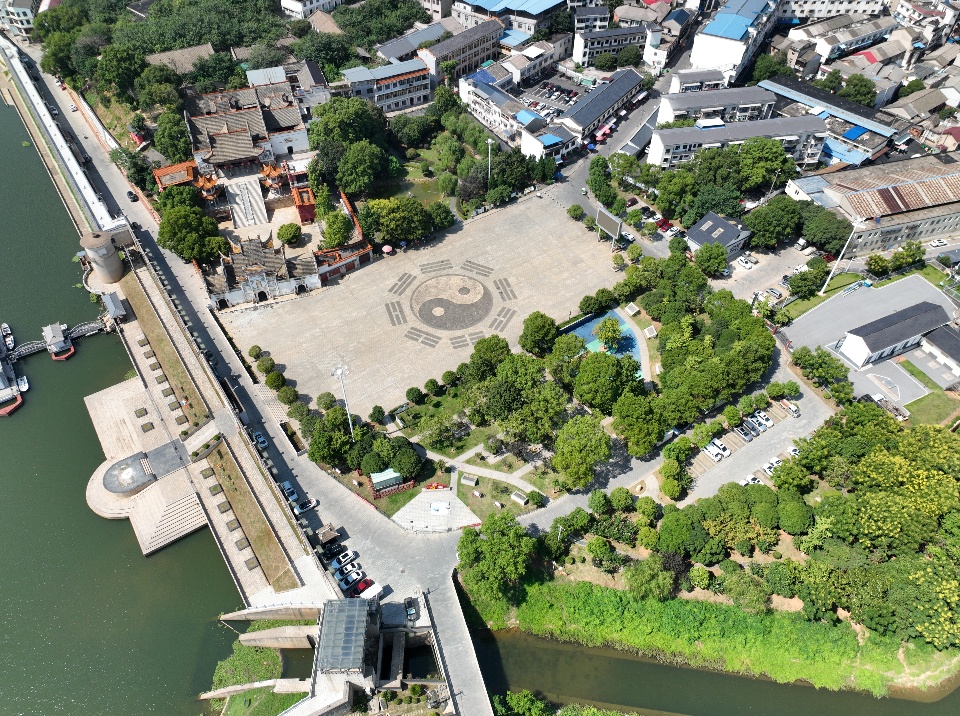}
    \end{subfigure}
    \hfill
    \begin{subfigure}[b]{0.195\textwidth}
        \includegraphics[width=\textwidth]{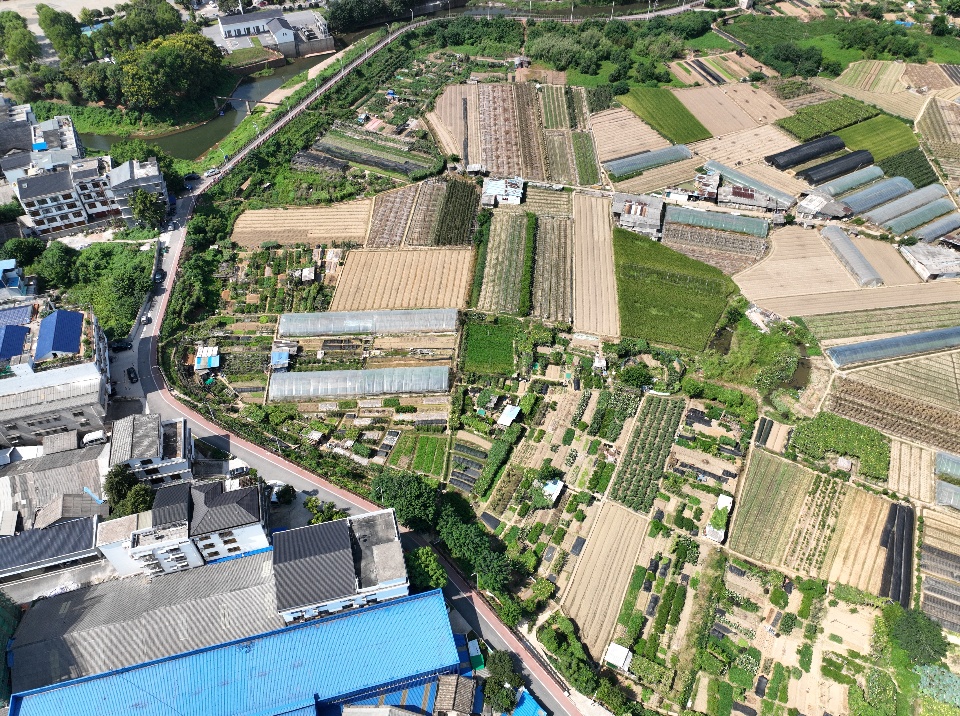}
    \end{subfigure}
    \hfill
    \begin{subfigure}[b]{0.195\textwidth}
        \includegraphics[width=\textwidth]{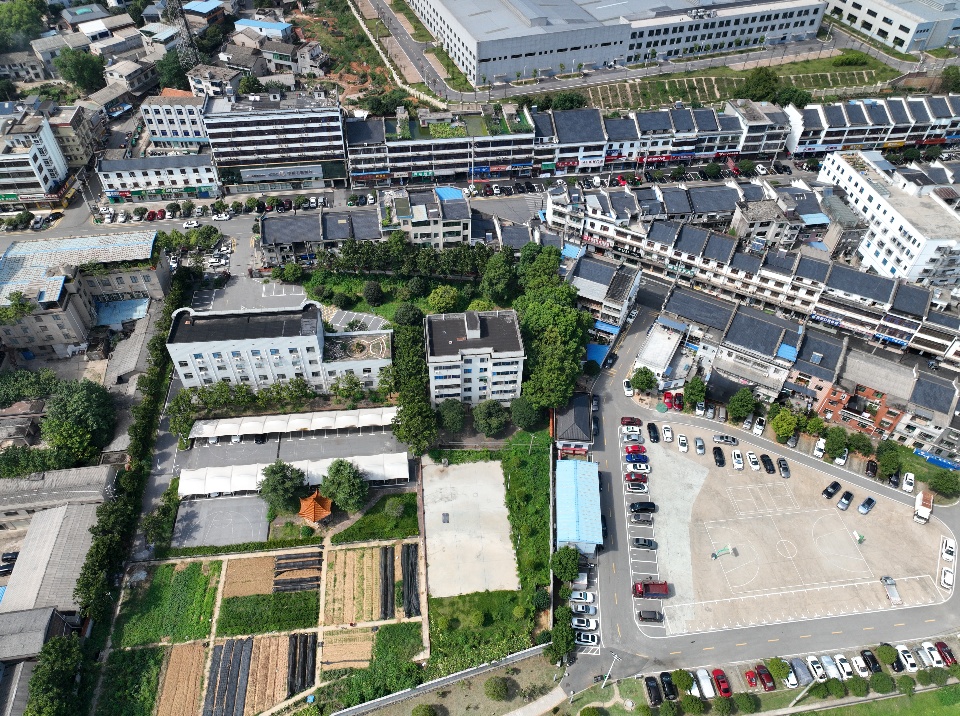}
    \end{subfigure}
    \hfill
    \begin{subfigure}[b]{0.195\textwidth}
        \includegraphics[width=\textwidth]{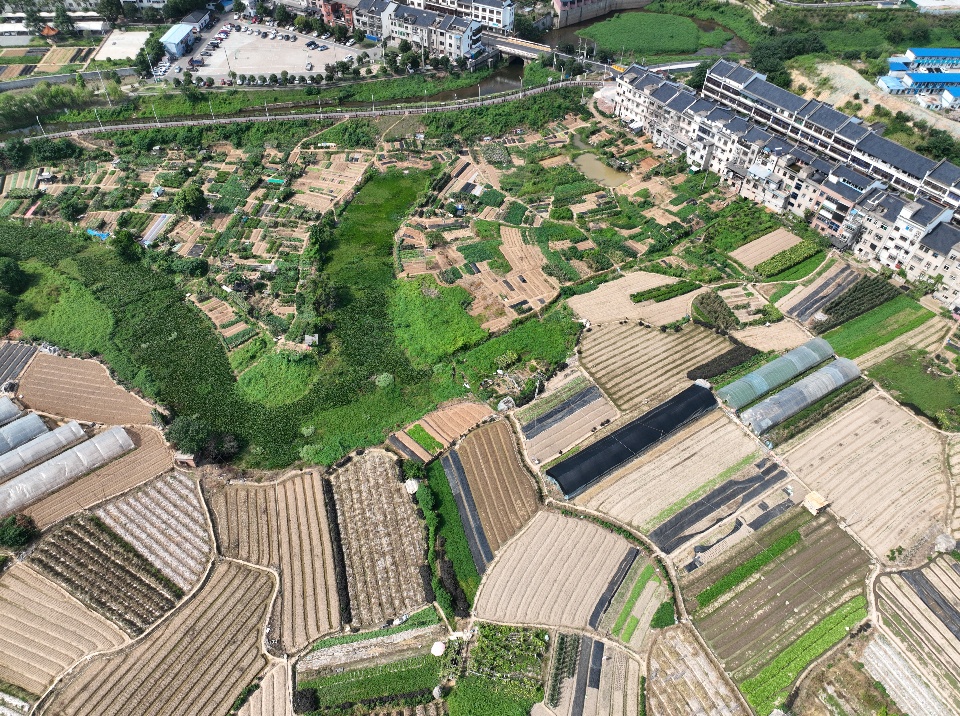}
    \end{subfigure}
    \hfill
    \begin{subfigure}[b]{0.195\textwidth}
        \includegraphics[width=\textwidth]{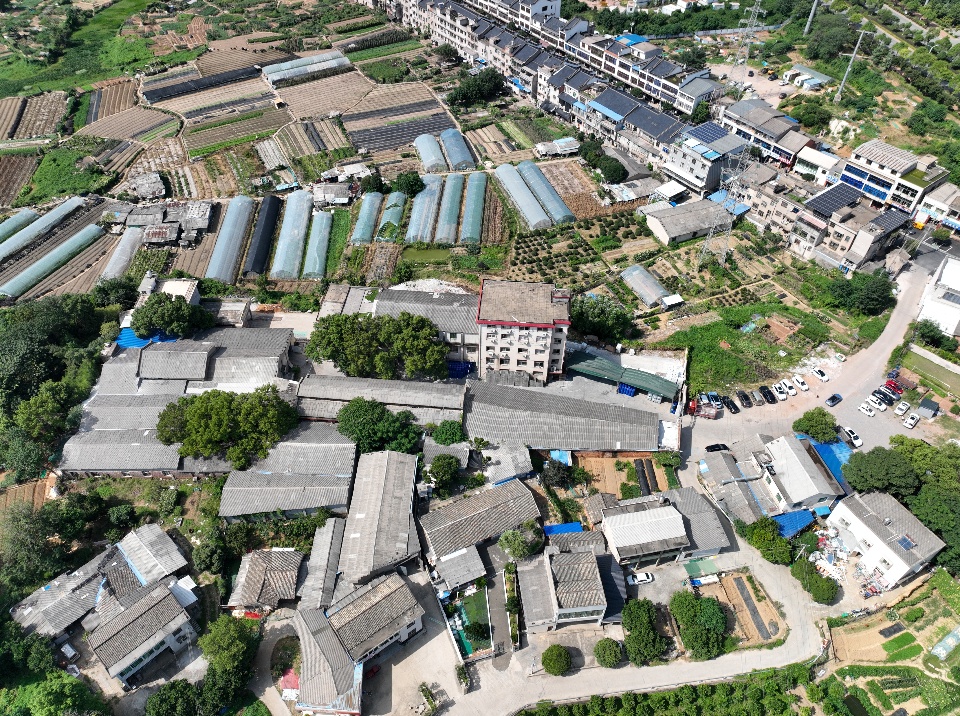}
    \end{subfigure}

    \begin{subfigure}[b]{0.195\textwidth}
        \includegraphics[width=\textwidth]{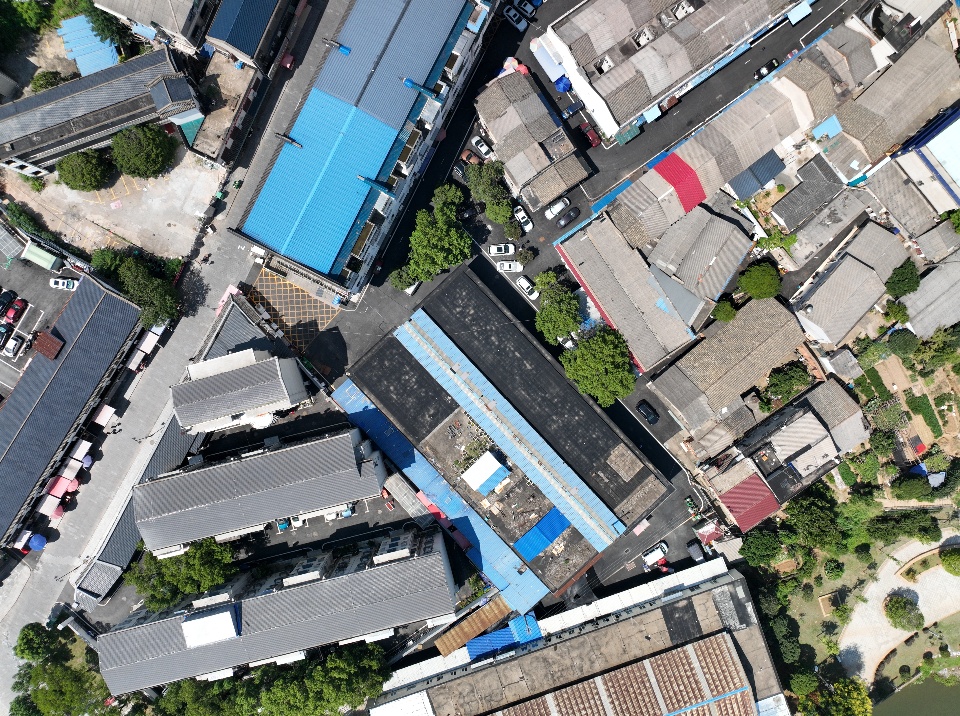}
    \end{subfigure}
    \hfill
    \begin{subfigure}[b]{0.195\textwidth}
        \includegraphics[width=\textwidth]{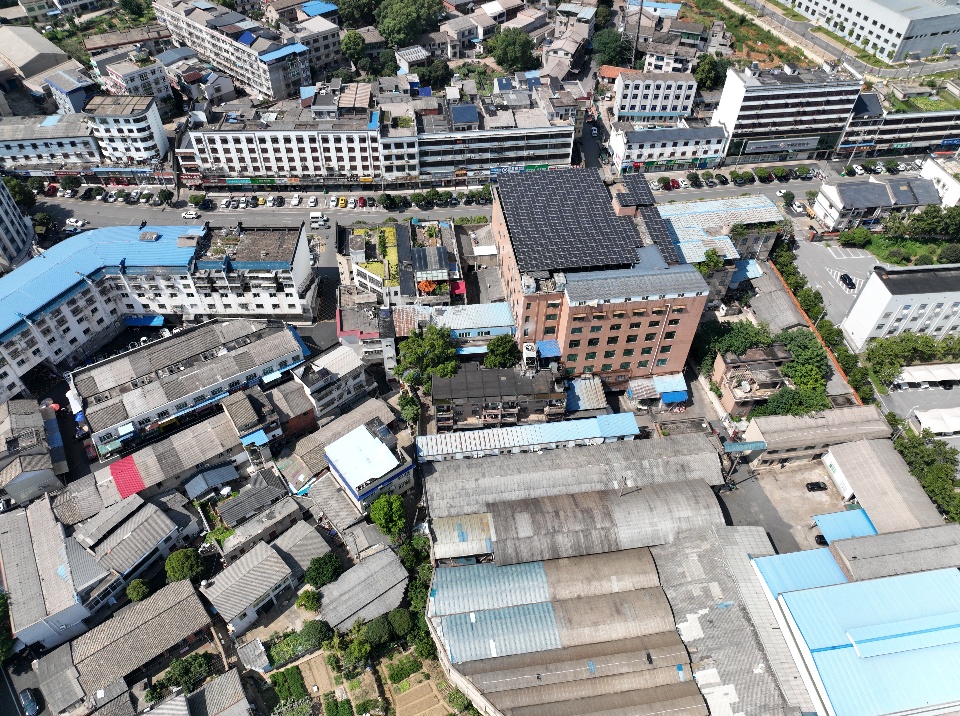}
    \end{subfigure}
    \hfill
    \begin{subfigure}[b]{0.195\textwidth}
        \includegraphics[width=\textwidth]{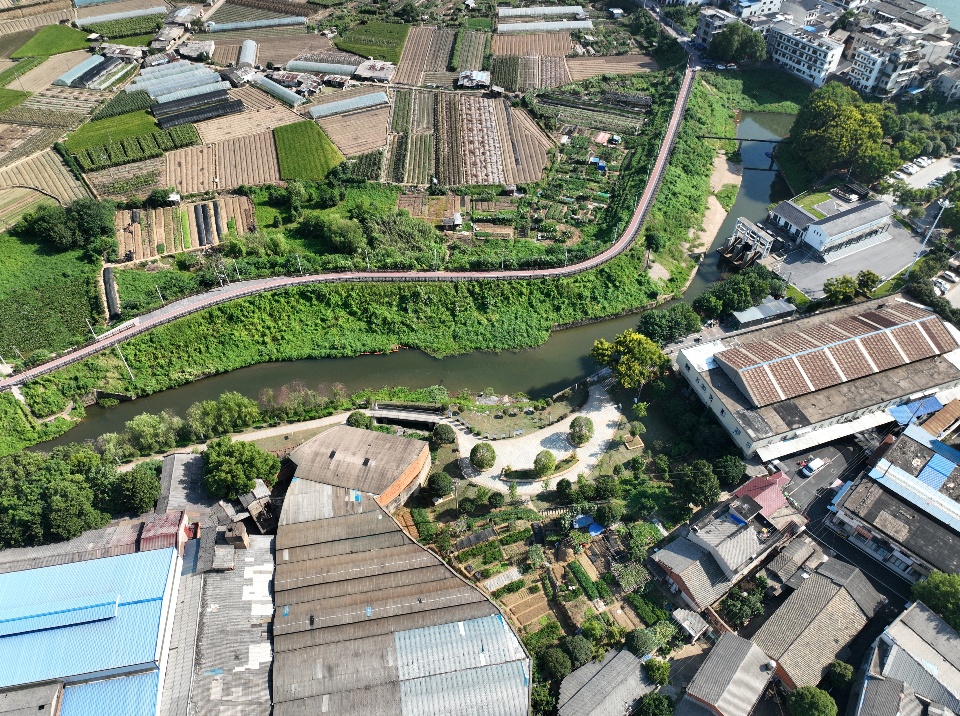}
    \end{subfigure}
    \hfill
    \begin{subfigure}[b]{0.195\textwidth}
        \includegraphics[width=\textwidth]{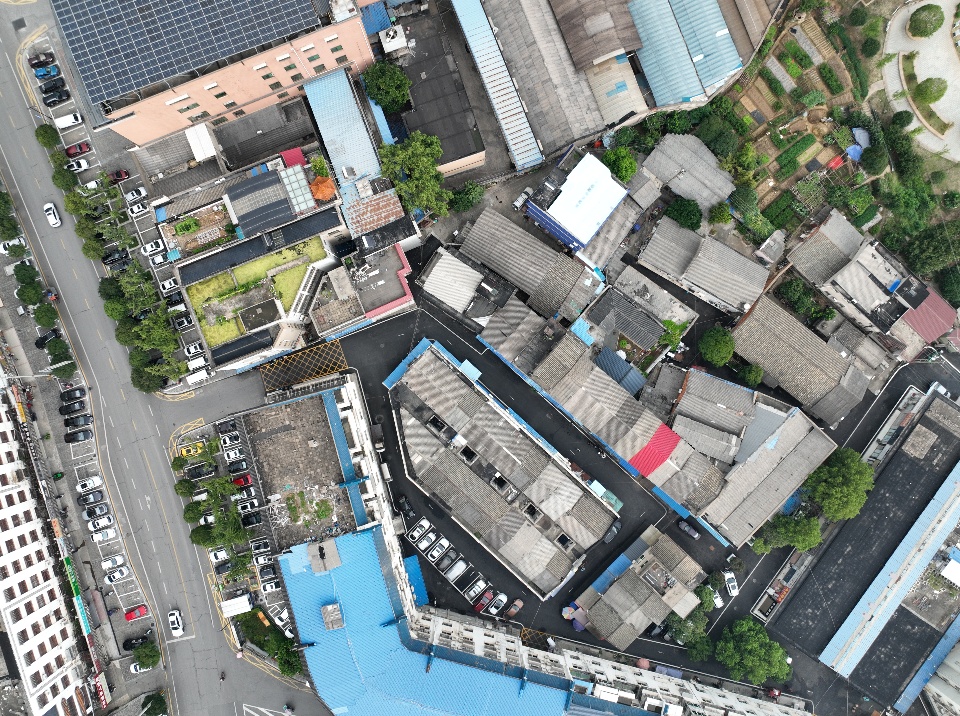}
    \end{subfigure}
    \hfill
    \begin{subfigure}[b]{0.195\textwidth}
        \includegraphics[width=\textwidth]{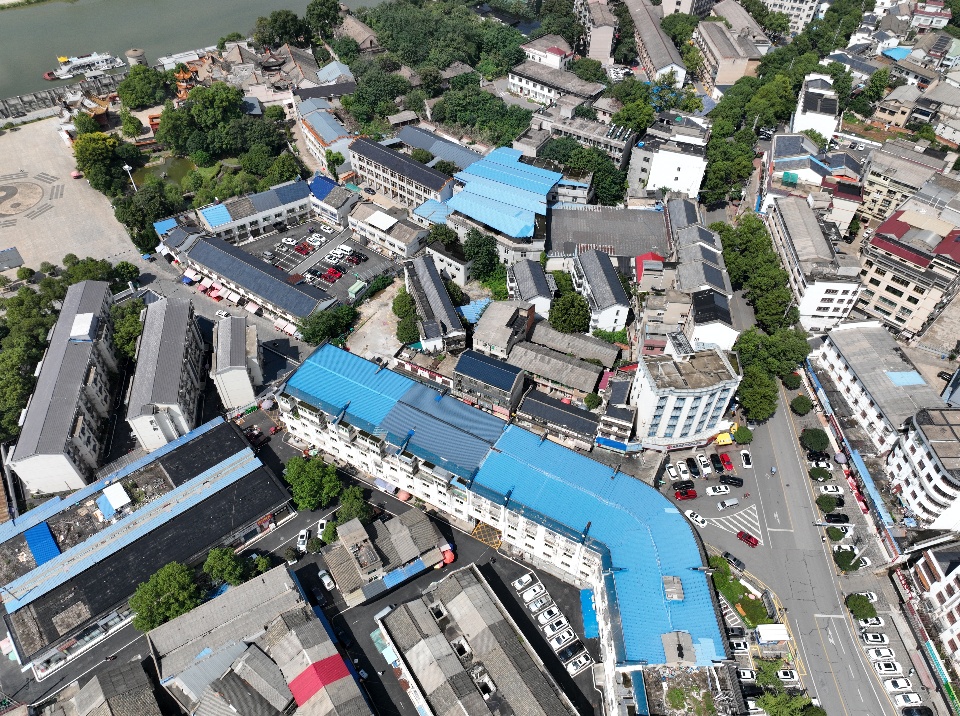}
    \end{subfigure}

    \begin{subfigure}[b]{0.195\textwidth}
        \includegraphics[width=\textwidth]{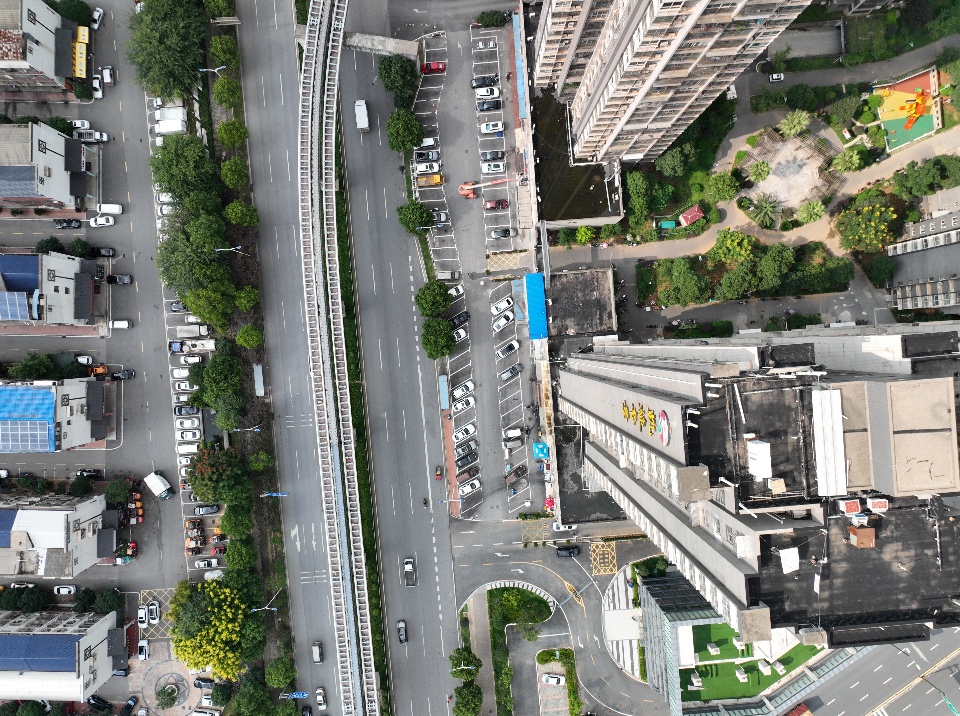}
    \end{subfigure}
    \hfill
    \begin{subfigure}[b]{0.195\textwidth}
        \includegraphics[width=\textwidth]{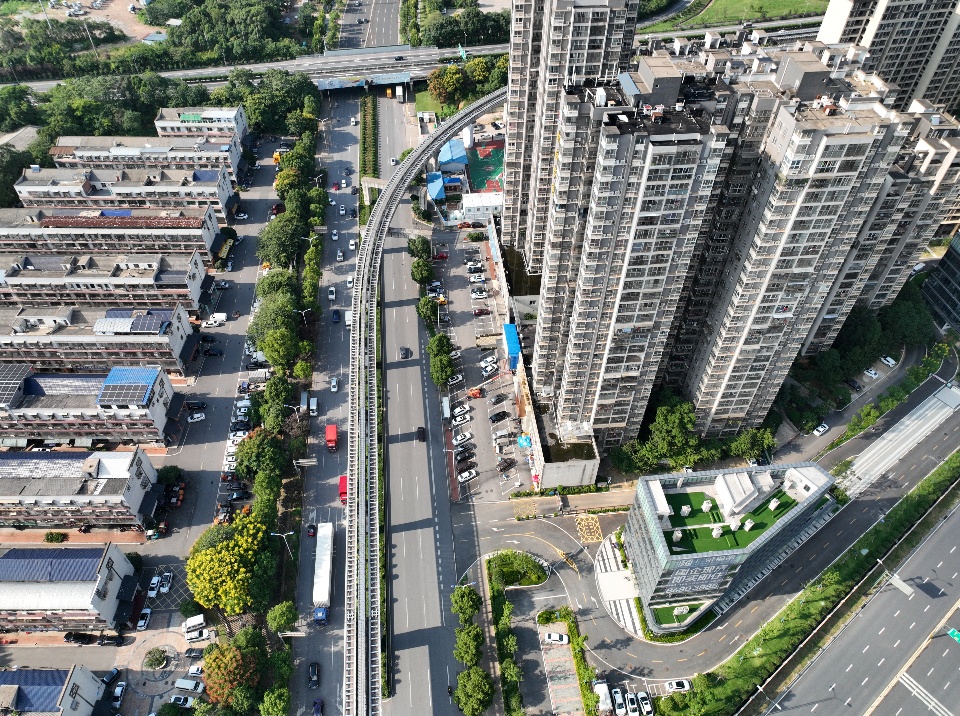}
    \end{subfigure}
    \hfill
    \begin{subfigure}[b]{0.195\textwidth}
        \includegraphics[width=\textwidth]{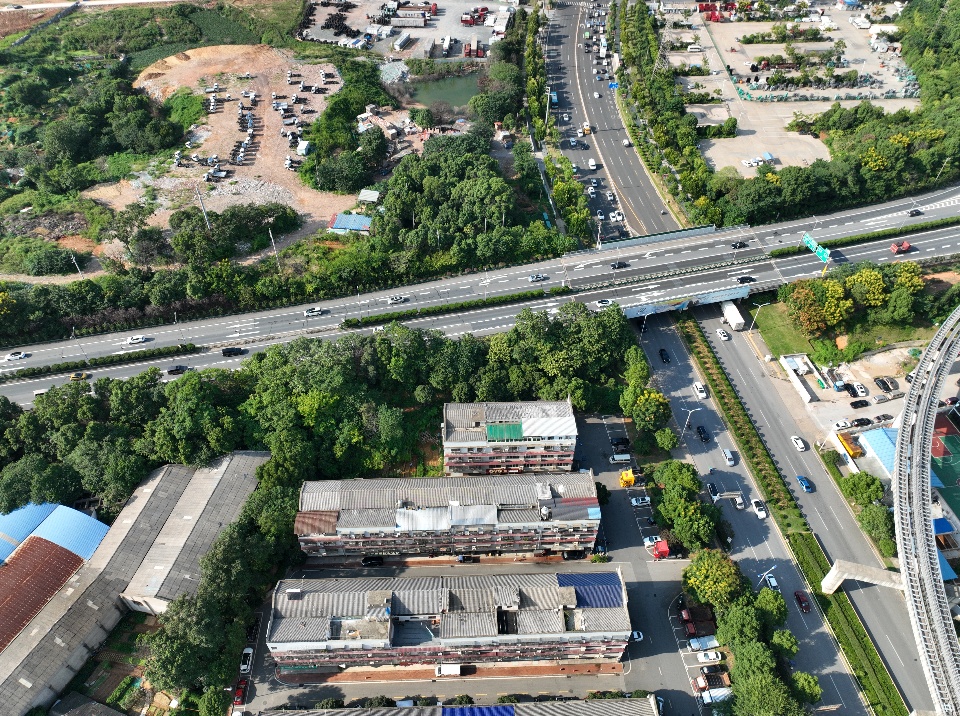}
    \end{subfigure}
    \hfill
    \begin{subfigure}[b]{0.195\textwidth}
        \includegraphics[width=\textwidth]{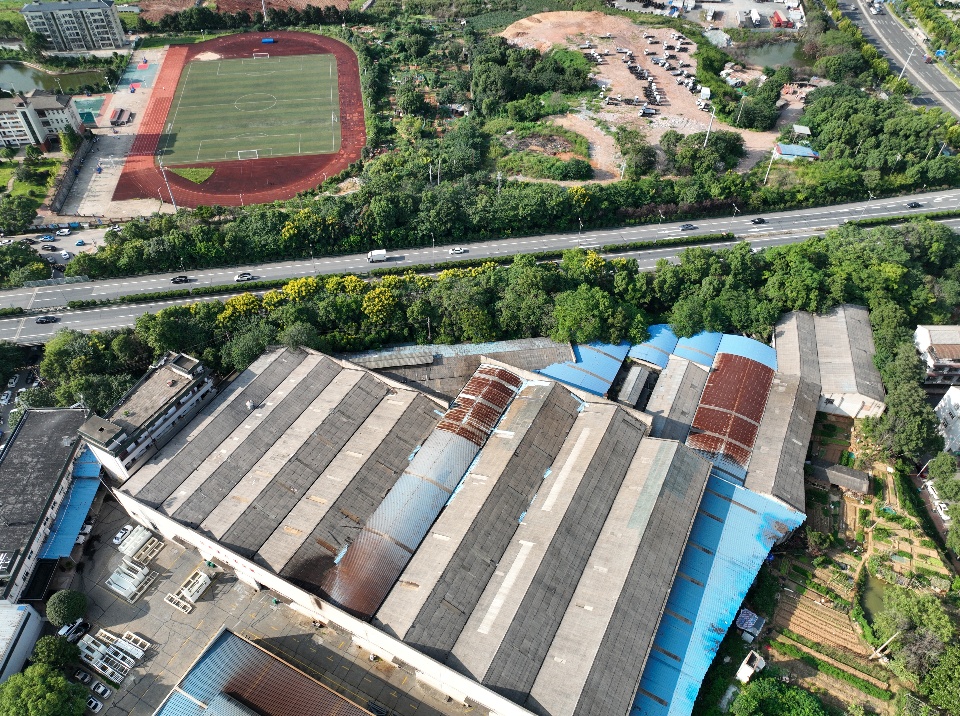}
    \end{subfigure}
    \hfill
    \begin{subfigure}[b]{0.195\textwidth}
        \includegraphics[width=\textwidth]{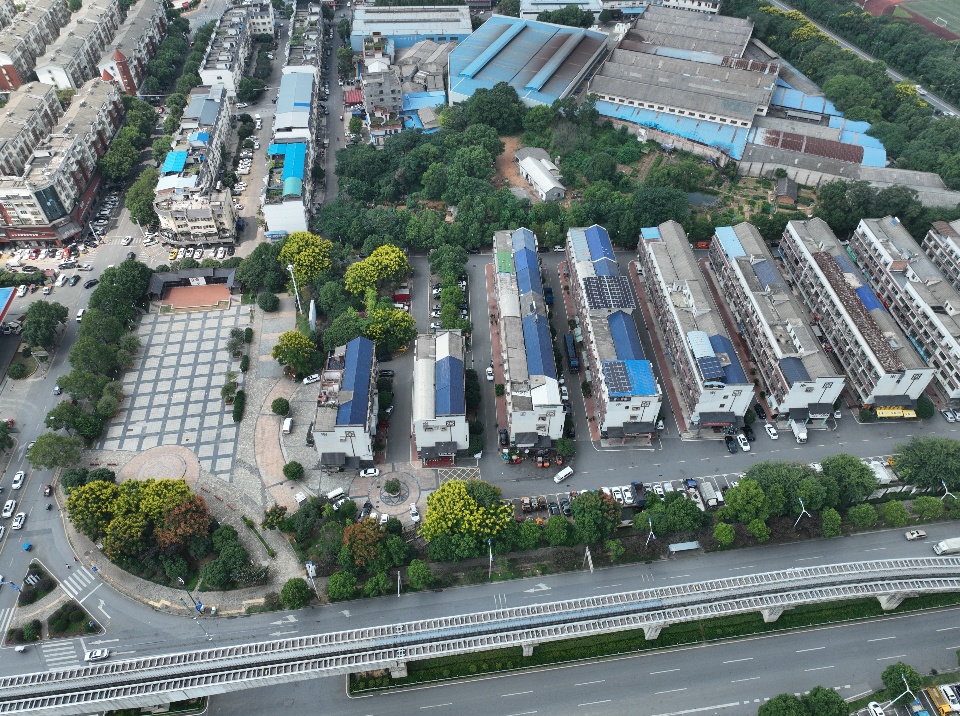}
    \end{subfigure}
    
    \caption{
        Example images from CS-Drone3D dataset. The first row shows the \textit{field} scene, the second row shows the \textit{hotel} scene, and the last row shows the \textit{bridge} scene. The images are acquired by a DJI drone at an altitude of about 140 meters.
    }
    \label{fig:my_figure}
\end{figure*}

\section{Description of CS-Drone3D Dataset}
CS-Drone3D is an oblique aerial dataset with a large spatial span, acquired by a DJI drone at an altitude of about 140 meters. It contains three scenes: field with 258 images, hotel with 197 images, and bridge with 275 images. The camera pose and intrinsic parameters are estimated by commercial software, and the images are undistorted as a pinhole camera model. The resolution is downsampled to 960x716 for easy distribution. Some examples are shown in Fig.~\ref{fig:my_figure}.

\section{Post-optimization for Regist3R}

Although Regist3R can achieve inference-only scene reconstruction, we can still design a post-optimization strategy to further improve its reconstruction accuracy. Here we propose a post-optimization scheme to ensemble multiple groups of camera poses from multiple reconstruction sequences.

In this scheme, multiple camera poses are ensembled according to their position in the spanning tree. In detail, we first execute Regist3R inference for $K$ times to obtain $K$ groups of camera poses and the layer at which the view is located in the spanning tree  $\{R_{k,i}, t_{k,i}, d_{k,i}\}_{k,i=1,1}^{K, N}$, where $N$ is the number of images. The goal is to find an optimal global camera poses $\{R_i, t_i\}_{i=1}^{N}$ that most matches all $K$ groups of poses under the sequence transformation $\{R'_k, t'_k, s_k\}_{k=1}^K$. From another perspective, we decompose the $K\times N$ camera transformations into $K$ sequence transformations and $N$ global camera transformations. The optimization target is formulated as:

\begin{equation}
  \min_{\substack{\{R_i, t_i\}\\\{R'_k, t'_k, s_k\}}}\sum_{k=1}^{K}\sum_{i=1}^{N}w_{k,i}\left[d_R(R_i, R'_kR_{k,i}) + \left\Vert t_i - (s_kR'_kt_{k,i}+t'_k) \right\Vert^2\right],
\end{equation}
where $w_{k,i}=w(d_{k,i})$ is the weight relative to the depth of view in spanning tree. The weight gets smaller when the view goes deeper. We heuristically define the weight as:
\begin{equation}
  w(d) = \exp\left(\frac{-5d}{\max_{k,i}(d_{k,i})}\right).
\end{equation}
$d_R(\cdot, \cdot)$ is the metric of rotation difference, such as the square of the angle difference in Lie algebra:
\begin{equation}
  d_R(R_1,R_2) = \left\Vert\log(R_1^TR_2)\right\Vert^2.
\end{equation}

The roots of spanning tree need to be distinct. In practice, we adopt K-medoids~\cite{schubert2019faster} on similarity matrix to select the roots. 

We evaluate the performance on DTU~\cite{aanaes2016large}, NRGBD~\cite{azinovic2022neural}, 7 scenes \cite{shotton2013scene} and CS-Drone3D, the results are reported in Tab.~\ref{tab:results}. We ensemble $K=3$ sequences for all datasets. In most of scenes, the pose precision consistently improved after ensemble, showing the effectiveness. But \textit{hotel} scene in CS-Drone3D dataset is an exception, as its inherent complexity leads to significant variations in reconstruction results when selecting different root nodes for initialization. This instability ultimately causes the ensemble-based optimization to fail.

\begin{table*}[htbp]
  \centering
  \caption{Ablation on post-optimization. The results are evaluated on DTU, NRGBD, 7 Scenes and CS-Drone3D dataset. Regist3R with post-optimization is denoted with Regist3R* in the table. We ensemble $K=3$ sequences for all datasets.}
  \label{tab:results}
  \begin{tabular}{l l *{3}{c} *{3}{c} c}
  \toprule
  Dataset & Method & \multicolumn{3}{c}{RRA} & \multicolumn{3}{c}{RTA} & mAA@30 \\
  \cmidrule(lr){3-5} \cmidrule(lr){6-8}
   & & @5 & @10 & @15 & @5 & @10 & @15 &  \\
  \midrule

  \multirow{2}{*}{DTU} 
  & Regist3R & 0.6790 & 0.9335 & 0.9697 & 0.6303 & 0.8712 & 0.9388 & 0.7988 \\ 
  & Regist3R* & \textbf{0.7483} & \textbf{0.9452} & \textbf{0.9738} & \textbf{0.7095} & \textbf{0.9078} & \textbf{0.9566} & \textbf{0.8292} \\
  \addlinespace[0.5em]

  \multirow{2}{*}{NRGBD} 
  & Regist3R & 0.8380 & 0.9998 & \textbf{1.0000} & 0.6656 & 0.8786 & 0.9458 & 0.8259 \\
  & Regist3R* & \textbf{0.8818} & \textbf{1.0000} & \textbf{1.0000} & \textbf{0.7379} & \textbf{0.9238} & \textbf{0.9666} & \textbf{0.8595} \\
  \addlinespace[0.5em]

  \multirow{2}{*}{7Scenes} 
  & Regist3R & 0.5730 & 0.9051 & 0.9661 & 0.4475 & 0.7053 & 0.8070 & 0.6771 \\
  & Regist3R* & \textbf{0.6314} & \textbf{0.9415} & \textbf{0.9889} & \textbf{0.5146} & \textbf{0.7806} & \textbf{0.8662} & \textbf{0.7313} \\
  \midrule

  \multicolumn{9}{l}{\textit{CS-Drone3D Dataset Scene Breakdown}} \\
  \midrule

  \multirow{2}{*}{Field} 
  & Regist3R & 0.7815 & \textbf{1.0000} & \textbf{1.0000} & 0.8964 & 0.9922 & 0.9980 & 0.8764 \\ 
  & Regist3R* & \textbf{0.9821} & \textbf{1.0000} & \textbf{1.0000} & \textbf{0.9766} & \textbf{0.9980} & \textbf{0.9996} & \textbf{0.9256} \\
  \addlinespace[0.5em]

  \multirow{2}{*}{Hotel} 
  & Regist3R & \textbf{0.8959} & \textbf{0.9612} & \textbf{0.9697} & \textbf{0.6120} & \textbf{0.8481} & \textbf{0.9132} & \textbf{0.7991} \\ 
  & Regist3R* & 0.4055 & 0.5179 & 0.5312 & 0.2498 & 0.4843 & 0.6129 & 0.3868 \\
  \addlinespace[0.5em]

  \multirow{2}{*}{Bridge} 
  & Regist3R & 0.8309 & \textbf{0.9783} & \textbf{0.9783} & 0.7877 & 0.9274 & 0.9554 & 0.8423 \\ 
  & Regist3R* & \textbf{0.8429} & \textbf{0.9783} & \textbf{0.9783} & \textbf{0.8090} & \textbf{0.9439} & \textbf{0.9654} & \textbf{0.8501} \\
  \midrule

  \multirow{2}{*}{CS-Drone3D} 
  & Regist3R & \textbf{0.8361} & \textbf{0.9798} & \textbf{0.9827} & \textbf{0.7654} & \textbf{0.9226} & \textbf{0.9555} & \textbf{0.8393} \\ 
  & Regist3R* & 0.7435 & 0.8321 & 0.8365 & 0.6785 & 0.8087 & 0.8593 & 0.7208 \\
  
  \bottomrule
  \end{tabular}
\end{table*}

\begin{figure*}[htbp]
  \centering
  \begin{subfigure}[b]{0.49\textwidth}
      \includegraphics[width=\textwidth]{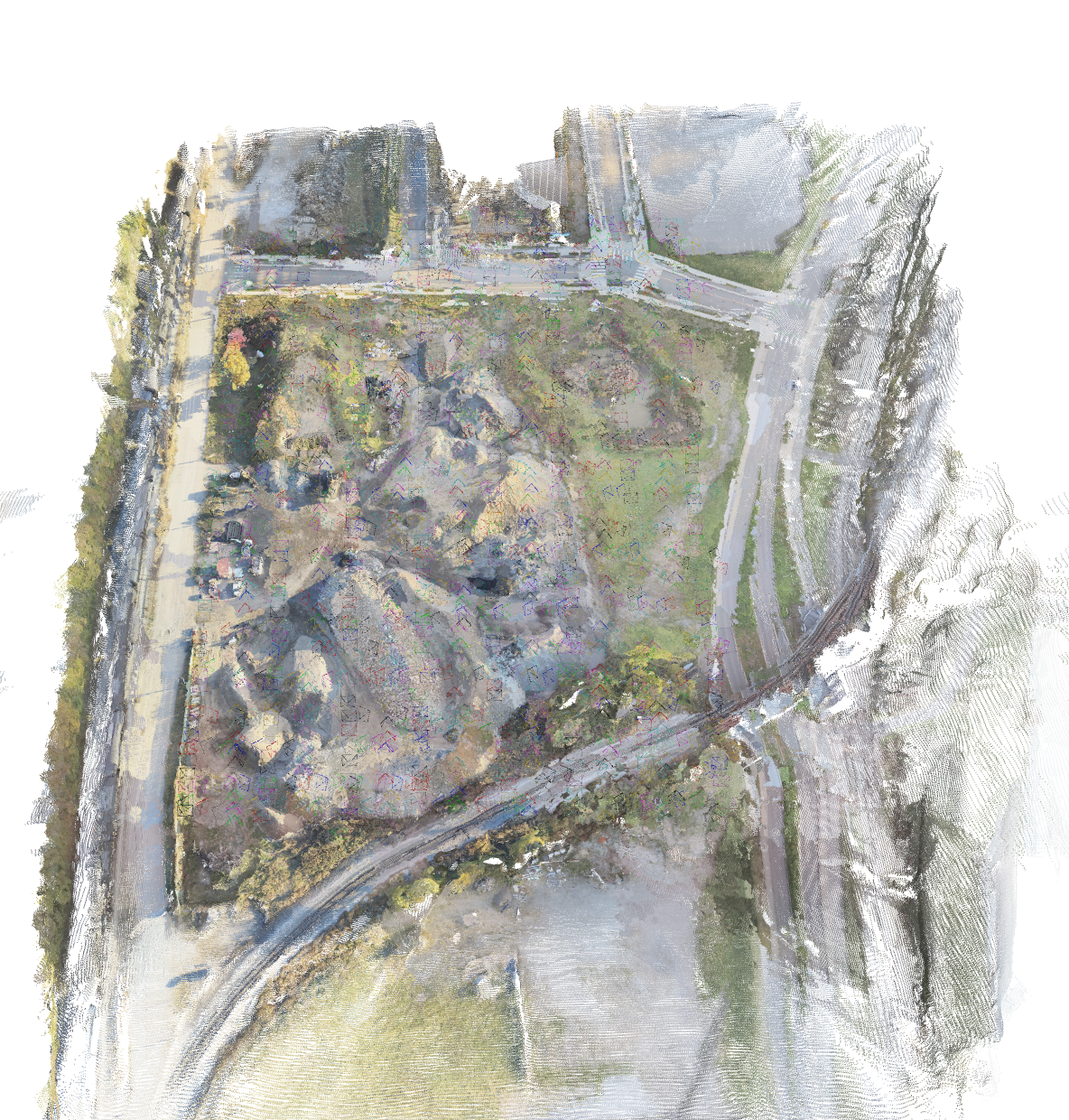}
      \caption{Cleaned Pointmaps}
      \label{fig:sub1}
  \end{subfigure}
  \hfill
  \begin{subfigure}[b]{0.49\textwidth}
      \includegraphics[width=\textwidth]{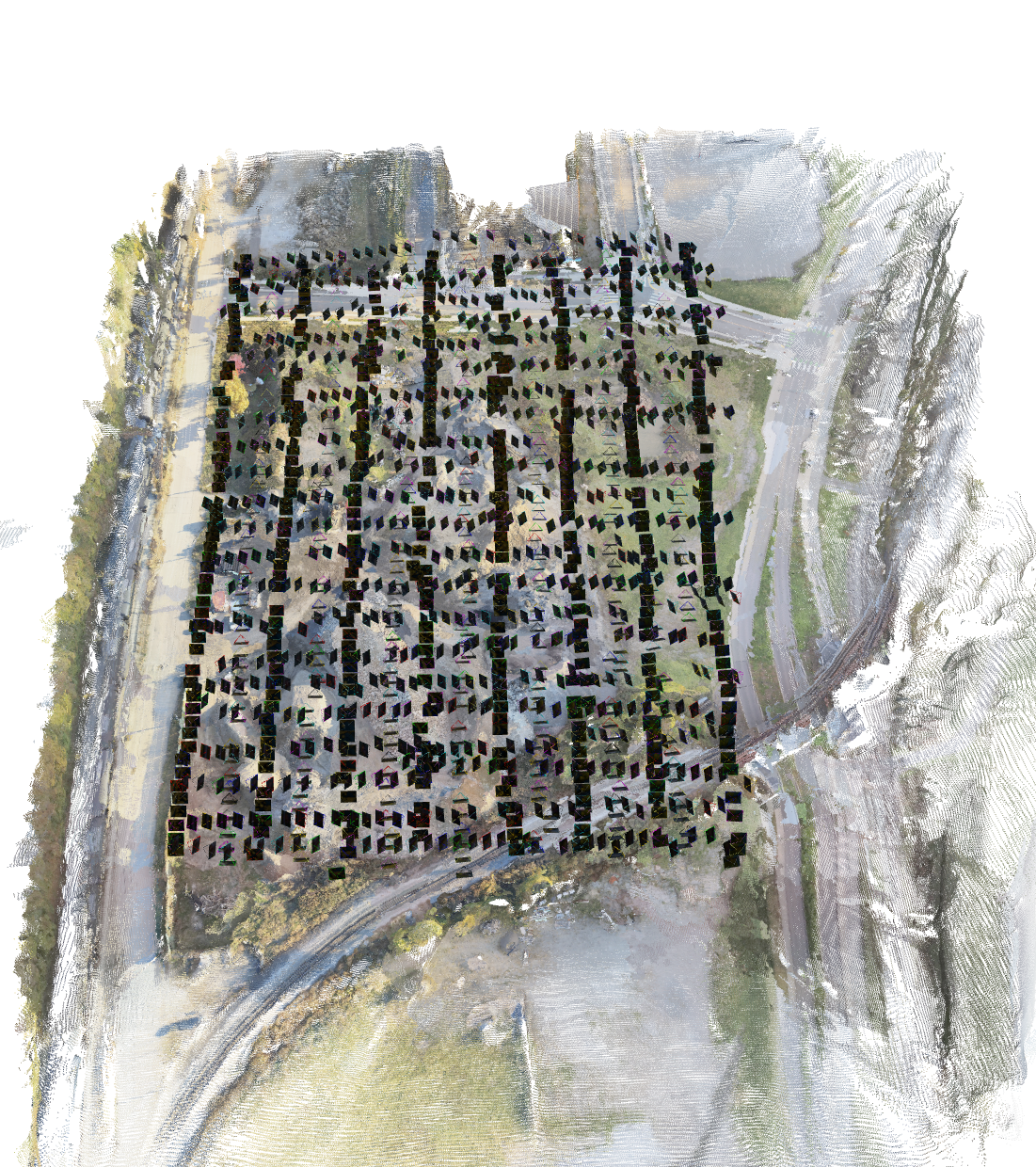}
      \caption{Camera poses}
      \label{fig:sub2}
  \end{subfigure}
  \begin{subfigure}[b]{0.49\textwidth}
      \includegraphics[width=\textwidth]{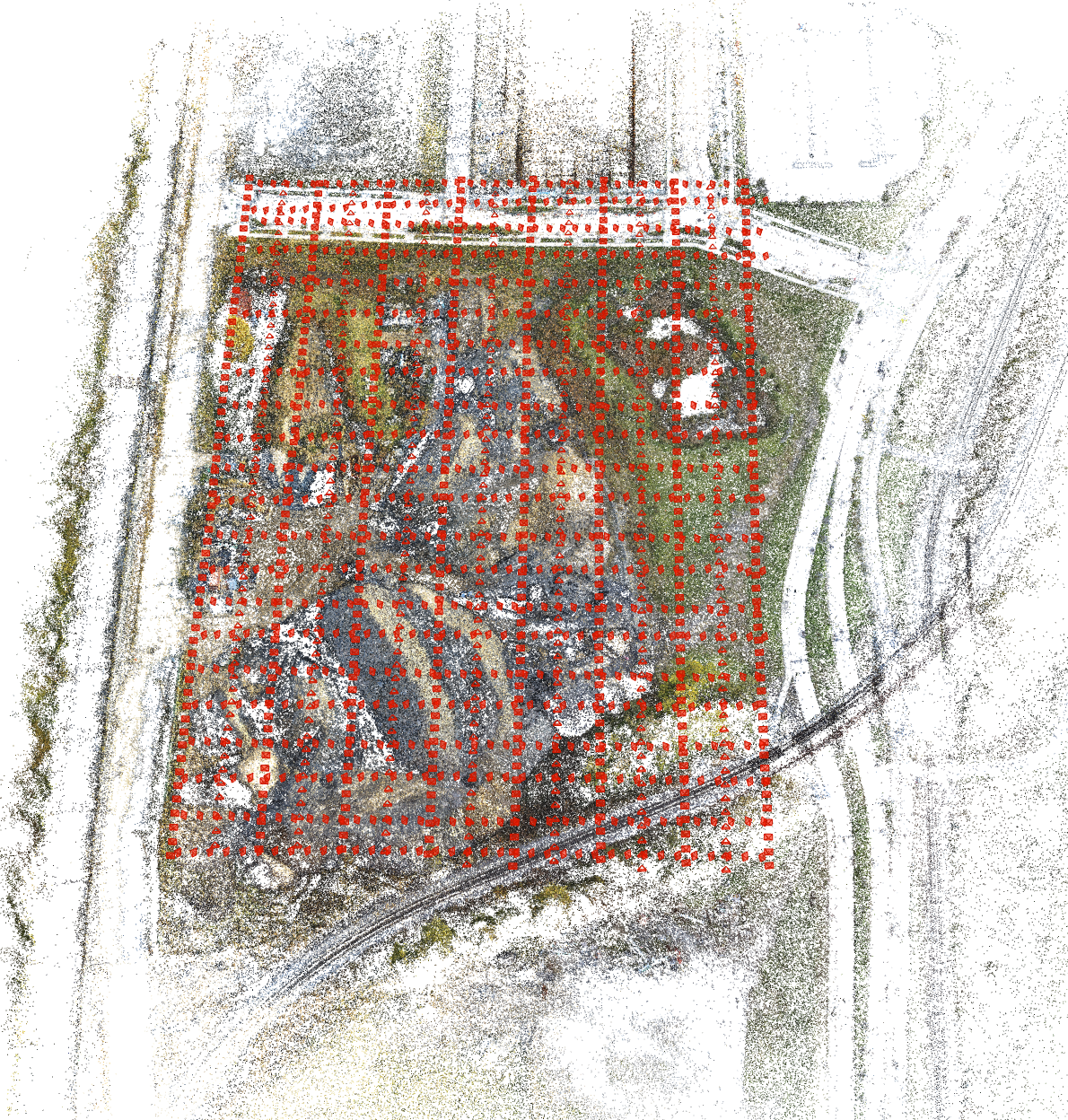}
      \caption{Ground truth poses and sparse points}
      \label{fig:sub3}
  \end{subfigure}
  \caption{Illustration of the reconstruction results of \textit{Rubble} scene from Hill19 dataset~\cite{turki2022mega}.}
  \label{fig:example}
\end{figure*}

\section{Reconstruction from 1000+ Views}

Scene reconstruction and camera pose estimation from large image collections is a long standing challenge. Traditional approaches like COLMAP~\cite{schonberger2016structure} often takes hours to days for reconstruction of collections containing thousands of views. Recent pointmap-based approaches achieves efficient reconstruction via model inference. But they still suffer from costly global alignment~\cite{wang2024dust3r,duisterhof2024mast3r}, model generalizability or high memory consumption~\cite{tang2024mv,yang2025fast3r,wang2025vggt}. Therefore previous pointmap-based approaches fail to reconstruct a scene with 1000+ views.
In this section, we demonstrate that Regist3R achieved this in a few minutes, taking \textit{Rubble} from Hill19 dataset~\cite{turki2022mega} as an example, which contains 1678 images, covering an area of about 1km$^2$.

To mitigate the impact of drift accumulation caused by excessive reconstruction sequence length on overall precision, we implement a hierarchical reconstruction strategy. This approach initially selects a limited number of key frames and employs a on-the-shelf multi-view model to predict their corresponding pointmaps, subsequently constructing minimum spanning tree (MST) forests with these key frames as root nodes. The remaining views are then registered using Regist3R based on the tree structure.

Specifically, we uniformly sample key frames at equidistant intervals and generate pointmaps for each with pretrained VGGT~\cite{wang2025vggt} model. These key frames constitute the root node set. During forest construction, we first introduce a virtual node with zero-distance connections to all root nodes. Crucially, we assign infinite distances to edges between root nodes to prevent their inclusion in the MST. Then an MST is constructed starting from the virtual node, which is subsequently removed to obtain the final MST forest. Finally, we perform Regist3R inference based on these spanning trees to register the remaining views. The entire reconstruction process is completed in approximately 5 minutes on one A100 GPU. An illustration of the reconstruction is shown in Fig.~\ref{fig:example}. 

\begin{figure}
    \centering
    \includegraphics[width=\linewidth]{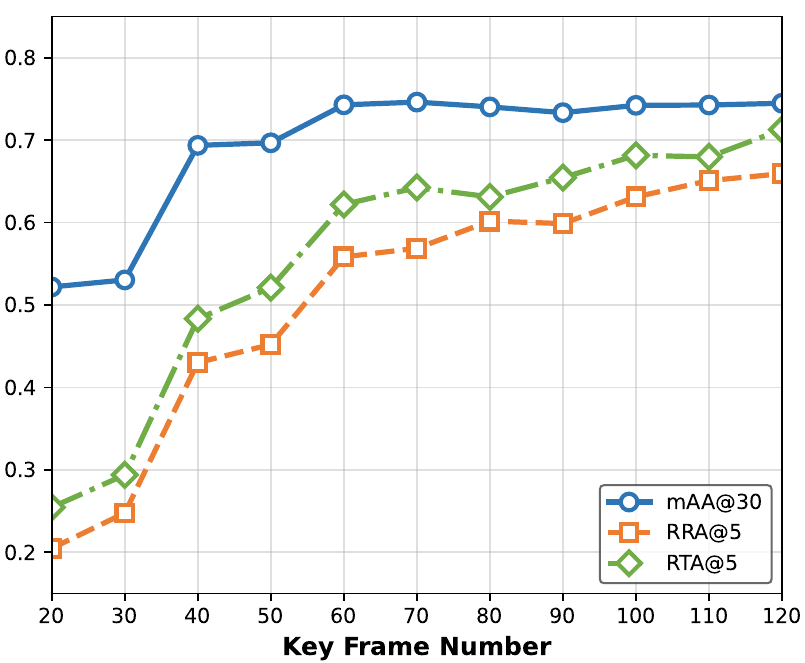}
    \caption{Number of keyframes versus camera pose accuracy.}
    \label{fig:rubble_pose_vs_anchor}
\end{figure}

We analysis the performance and efficiency of the proposed hierarchical pipeline in Fig.~\ref{fig:rubble_pose_vs_anchor} and Tab.~\ref{tab:extended_time}. As shown in Fig.~\ref{fig:rubble_pose_vs_anchor}, the camera pose accuracy is improved with the increase of keyframes. But the memory bound limits the number of key frames from further increasing. In Tab.~\ref{tab:extended_time}, we use 100 key frames to evaluate the time consumption of the pipeline. The pipeline takes about 5 minutes to finish the reconstruction, containing the time of model loading, image loading and preprocessing, ASMK-based image retrieval, model inference, and pose solving. The net inference time is 70 seconds on 1678 images (100 images for VGGT and 1578 images for Regist3R). The FPS is 7.36 for VGGT inference while 27.89 for Regist3R inference.

\begin{table}[h]
\centering
\caption{Total Time, Net Time, and FPS Analysis}
\label{tab:extended_time}
\begin{tabular}{lcccc}
\toprule
Step               & Total Time (s) & Net Time (s) & FPS    \\
\midrule
ASMK Retrieval     & 92.73          & --           & --     \\
VGGT Inference     & 31.59          & 13.59        & 7.36   \\
Regist3r Inference & 129.36         & 56.55        & 27.89  \\
Solve PnP          & 79.80          & --           & --     \\
\midrule
\textbf{Total}     & \textbf{333.48} & --           & --     \\
\bottomrule
\end{tabular}
\end{table}

\section{Confidence Visualization}

\begin{figure*}
    \includegraphics[clip,trim=5mm 5mm 5mm 5mm, width=\textwidth]{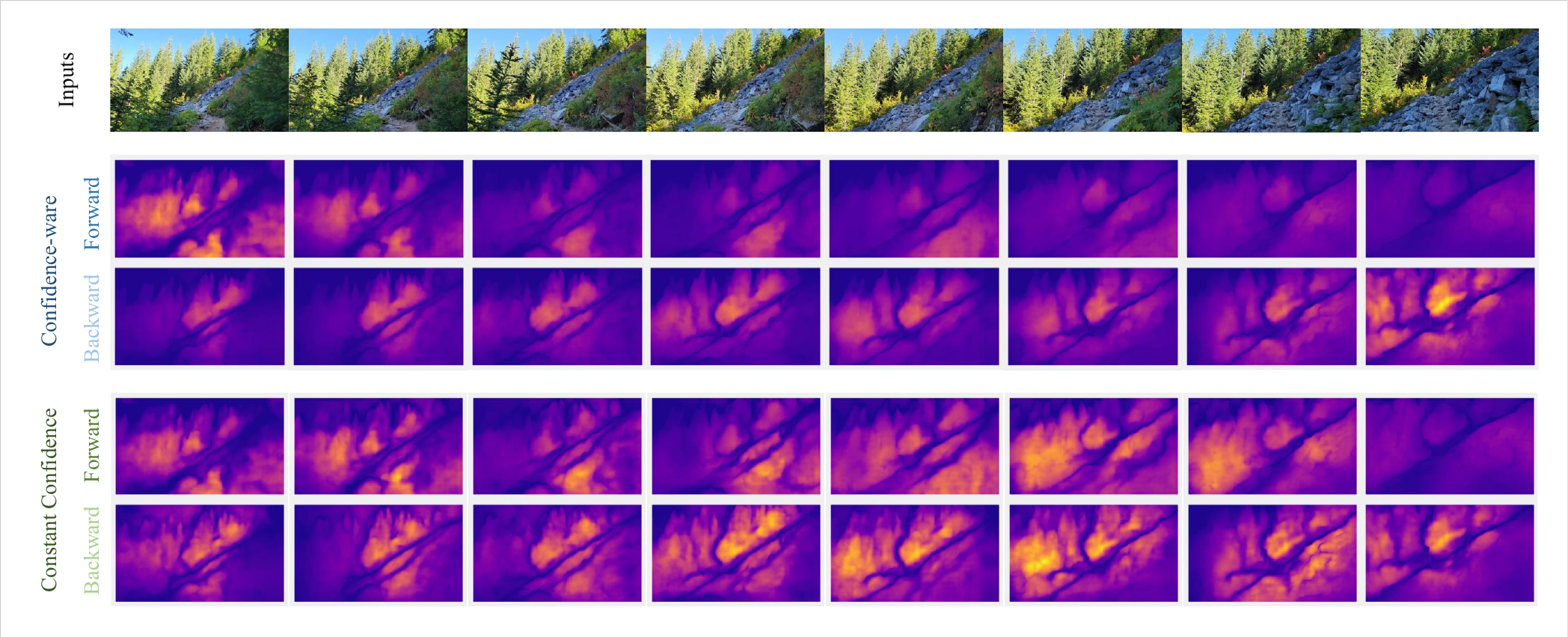}
    \caption{Visualization of the confidence decay. The first row shows the input images, the following two rows are the confidence heatmaps when the confidence-aware autoregression is enabled, with the first row showing the forward reconstruction sequence (from left to right) and the second row showing the backward (from right to left). The last two rows is the confidence heatmaps when the confidence-aware autoregression is disabled (input confidence is set to one). Obvious confidence decay can be observed when the autoregression is enabled, while the confidence is uniformly distributed when the autoregression is disabled.}
    \label{fig:confidence decay}
\end{figure*}



\begin{table*}[htbp]
\centering
\caption{Performance Comparison of Confidence Strategies}
\label{tab:confidence_comparison}
\footnotesize
\begin{tabular}{l*{7}{S[table-format=1.3]}}
\toprule
\multirow{2}{*}{\textbf{Strategy}} & 
\multicolumn{7}{c}{\textbf{Forward Sequence}} \\
\cmidrule(lr){2-8}
 & {mAA@30} & {RRA@5} & {RRA@10} & {RRA@15} & {RTA@5} & {RTA@10} & {RTA@15} \\
\midrule
\textbf{Confidence-aware} & 0.798 & 0.978 & 1.000 & 1.000 & 0.600 & 0.822 & 0.933 \\
Constant confidence & 0.728 & 0.822 & 1.000 & 1.000 & 0.444 & 0.756 & 0.844 \\
\cmidrule(r){1-8}
\textbf{Strategy} & 
\multicolumn{7}{c}{\textbf{Backward Sequence}} \\
\cmidrule(lr){2-8}
 & {mAA@30} & {RRA@5} & {RRA@10} & {RRA@15} & {RTA@5} & {RTA@10} & {RTA@15} \\
\midrule
Confidence-aware & 0.381 & 0.578 & 1.000 & 1.000 & 0.022 & 0.200 & 0.400 \\
\textbf{Constant confidence} & 0.558 & 1.000 & 1.000 & 1.000 & 0.044 & 0.378 & 0.689 \\
\bottomrule
\end{tabular}
\end{table*}

As shown in the ablation study, the confidence-aware autoregressive training improves the performance of Regist3R. This section quantitatively analyze it by visualizing the confidences, taking 10 frames from a video captured by hand-holded camera from LocalRF dataset~\cite{meuleman2023localrf} for evaluation, as shown in Fig.~\ref{fig:confidence decay}.

We perform a sequential reconstruction from the first frame to the last frame (forward) and vise versa (backward) under the existence (row 2-3) or absence (row 4-5) of confidence-aware autoregressive training. The confidence heatmaps are visualized. Obvious confidence decay can be observed when the autoregression is enabled. This shows that the model takes the confidence of the previous step into consideration when evaluating the pointmap confidence of the current step, which shows that the autoregressive training has played the expected role. As contrasted, the confidence is uniformly distributed when the autoregression is disabled. 

It is worth noting that confidence aware autoregressive training does not always improve performance, as shown in Tab.~\ref{tab:confidence_comparison}. The performance of the backward sequence becomes worse with confidence aware setting. One possible explanation is that during the backward reconstruction, some areas of the target view are not observed by the reference view, which reduces confidence and makes the model not dare to use this area as a reference to predict subsequent pointmaps, resulting in reduced pose accuracy. At this time, blind confidence may help the model to reconstruct the sequence more firmly.

\end{document}